%% file: 0-main.tex
\documentclass[fleqn,usenatbib]{mnras}

\usepackage[T1]{fontenc}

\DeclareRobustCommand{\VAN}[3]{#2}
\let\VANthebibliography\thebibliography
\def\thebibliography{\DeclareRobustCommand{\VAN}[3]{##3}\VANthebibliography}

\usepackage{graphicx}	% Including figure files
\usepackage{amsmath}	% Advanced maths commands
\usepackage{amssymb}	% Extra maths symbols

\usepackage{ulem}	% 20220203, for over-plot delete line.
\newcommand\Hb{H$\beta$}
\newcommand\Ha{H$\alpha$}
\newcommand\UpperRoman[1]{\uppercase\expandafter{\romannumeral#1}}      % specType by upper Roman number.
\newcommand\CaII{Ca \uppercase\expandafter{\romannumeral2}}             % Ca II

\title[Magnetic activity on HD 134319]
{
The investigation on Magnetic Activity of HD 134319 based on TESS Photometry and Ground-based Spectroscopy
}

\author[F. Xu et al.]{
Fukun Xu%
$^{\href{https://orcid.org/0000-0002-8618-3551}{\includegraphics[width=8px]{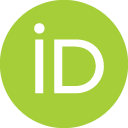}}}$,%
$^{1,2}$\thanks{Email: \href{mailto: xufukun@ynao.ac.cn}{xufukun@ynao.ac.cn}(FX); \href{mailto: shenghonggu@ynao.ac.cn}{shenghonggu@ynao.ac.cn}(SG)}
Shenghong Gu$^{1,2,3}$\footnote[1]{\href{mailto: shenghonggu@ynao.ac.cn}{shenghonggu@ynao.ac.cn}(SG)}
and
Panogiotis Ioannidis$^{4}$
\\
$^{1}$Yunnan Observatories, Chinese Academy of Sciences, Kunming 650216, China\\ %; shenghonggu@ynao.ac.cn \\
$^{2}$Key Laboratory for the Structure and Evolution of Celestial Objects, Chinese Academy of Sciences, Kunming 650216, China\\
$^{3}$School of Astronomy and Space Science, University of Chinese Academy of Sciences, Beijing 101408, China\\
$^{4}$Hamburger Sternwarte, Universit{\"a}t Hamburg, Gojenbergsweg 112, D-21029 Hamburg, Germany\\
}

\date{Accepted xxx. Received xxx; in original form xxx}

\pubyear{2022}

\begin{document}
\label{firstpage}
\pagerange{\pageref{firstpage}--\pageref{lastpage}}
\maketitle

\begin{abstract}
We present analysis of the starspot properties and chromospheric activity on HD 134319 using high precision photometry by TESS in sectors 14--16 (T1), 21--23 (T2) %in sectors 14--16 and 21--23 
and high-resolution spectroscopy by OHP/ELODIE and Keck/HIRES during years 1995--2013.
%
%Firstly rotation period of $P=4.436391\pm0.00137$ days was estimated using GLS method and used in following analysis.
% = Method
%Then the light curve was split to sliding chunks, each of which in 5 days length and was modelled by a two-spot model 
%By sliding % a 5 days window with step 1 day 
We applied a two-spot model with GLS determined period $P=4.436391\pm0.00137$ day to model chunks sliding over TESS light curve, % and chose the best-fit solution by assuming that adjacent chunks hold similar configurations. 
and measured relative equivalent widths of Ca II H and K, H$\beta$ and H$\alpha$ emissions by subtracting overall spectrum from individual spectrum.
% = Result
% - configuration
It was found that a two-spot configuration, i.e. a primary, slowly evolving and long-lasting spot (P) plus a secondary and rapidly evolving spot S, %\textcolor{red}{\sout{was sufficient for spot modelling}}
was capable of explaining the data, although the actual starspot distribution is unable to derived from collected data.
Despite the spot radius-latitude degeneracy revealed in the best-fit solutions, a sudden alternation between P and S radii followed by gradual decrease of S in T1 and the decrease of both P and S from T1 to T2 were significant, corresponding to the evolution of magnetic activity.
Besides, S revealed rotation and oscillatory longitude migration synchronized to P in T1, but held much larger migration than P in T2. This might indicate the evolution of internal magnetic configuration.
% contemporary 
Chromospheric activity indicators were found to tightly correlated with each other and revealed rotational modulation as well as a long-term decrease of emissions, implying the existence and evolution of magnetic acitivity on HD 134319.
%\textcolor{red}{\sout{Their rotational modulation around year 1996 coincided well with photometric variations at epoch 1995.38 and indicated also two active regions, implying that such a configuration might be frequently emerged on HD 134319.}}
%
\end{abstract}


%%%%%%%%%%%%%%%%%%%%%%%%%%%%%%%%%%%%%%%%%%%%%%%%%%
\input{1-intro}
\input{2-observation-reduction}

\input{3-model}
\input{4-result-discussion}
\input{5-conclusion}
\input{6_Acknowledgement}
\bibliographystyle{mnras}
\bibliography{0-main}

%%%%%%%%%%%%%%%%%%%%%%%%%%%%%%%%%%%%%%%%%%%%%%%%%%

% Don't change these lines
\bsp	% typesetting comment
\label{lastpage}
\end{document}

%% file: 1-intro.tex
\section{Introduction}
%% ======
%% Magnetic activity.
Stellar magnetic activity arising from the enhancement of magnetic flux generated at the tachocline, penetrated the outer convection zone and then exceeded the surface of late-type stars manifests itself in abundant phenomena, such as starspots, plages and flares, causing series of distortions detectable in both photometric and spectroscopic observations. 
Periodic or quasi-periodic variations in photometric measurements of cool stars are widely attributed to the rotational modulation of starspots. This makes starspot a good tracer of stellar rotation and detector of differential rotation which plays a critical role in generating and maintaining the solar-like magnetic activity through an $\alpha\Omega$ dynamo \citep{Isik2011}.
Meanwhile, emissions of chromospheric activity indicators such as the Ca II H \& K resonance and Balmer lines are generally used as diagnoses of magnetic activity in stellar outer layers and also show modulation under stellar rotation \citep{Vida2015}.

In analogy to the case of the Sun, %'s 11-yr cycle revealed by sunspots activity, 
the magnetic activity on active stars is believed to reveal both spatially inhomogeneous distribution and temporally evolution.
The former creates inhomogeneous configuration of the magnetic field accompanying by the emergence of photospheric starspot and chromospheric excitation \citep{Hempelmann2016, Balona2019}, while the latter exhibits as evolution of starspot in terms of emerging, migrating, decreasing and vanishing during its life %and their variable contrast and proportion to the photosphere, 
as well as the variation of chromospheric proxies in multiple timescales \citep{Mittag2019}.
%Meanwhile, as magnetic activity can potentially put feedback on the dynamical process under rotation, the DR can also evolve with time in principle due to exchange between magnetic and kinetic energy \citep{Petit2002}.
%Although this was reported only on AB Dor \citep[][see]{CollierCameron2002, Donati2003, Jeffers2007} and discussed by a few investigations \citep[][e.g.]{2019MNRAS.489.5556Y, Xu2021}, the evolution of starspots was notable on cool stars and studied by a bulk of studies \citep[e.g.][]{Henry1995, Hussain2002, Lanza2006, Xiang2014}.
By long-term monitoring on photometric and spectroscopic variations, one can derive the starspot distribution, evolution and its connection with the chromospheric activity, which were reported on several targets and believed to be widely existed on late-type stars \citep[e.g.][]{Vida2015, Morris2018, Xu2021}.
However such a study is usually non-trivial due to the limit in acquiring high precision data of target in distance.

%TESS
In recent years, space-borne telescopes such as MOST \citep{Walker2003}, CoRoT \citep{Auvergne2009}, Kepler \citep{Koch2010} and most recent Transiting Exoplanet Survey Satellite \citep[TESS,][]{Ricker2015} produced bulk of long-term photometric data.
Since its initiation, TESS was designed to cover $>80\%$ of the whole sky and deliver light curves (LCs) spanning from $27$ days to years for millions of stars, providing us the opportunity to analyse the starspot configuration and its evolution on stellar photosphere with high-precision \citep{Reinhold2013, Namekata2020}.

%% basic info of GJ 577.
The star HD 134319 (TIC 202426247, GJ 577, Gl 577, HIP 73869, with magnitude $V = 8.41$ mag, $G = 8.23$, and color $B-V = +0.677$), is a young \citep[$\sim 0.625\,\mathrm{Gyr}$,][]{Montes2001, McCarthy2001, Decin2003} active G5 main-sequence star of BY Dra type \citep{Soderblom1985} which can be taken as a young solar analogue with a shorter period of about $4.448\,\mathrm{day}$ \citep{Messina1998}.
Locating in the north semi-sphere, was observed in six sectors by TESS (14--16 and 21--23) spanning over $300$ days, revealing significant active regions surviving over 90 days as well as evidences of evolution in both short and long timescales.
%By joint analysing with %spectroscopic archive data observed by the high-resolution echelle spectrographs, the HIRES spectrometer on the Keck \UpperRoman{1}-10m telescope of Mount Wilson Observatory \citep[Keck/HIRES,][]{Vogt1994} and the ELODIE on the 1.93m telescope of Observatoire de Haute-Provence \citep[OHP/ELODIE,][]{ELODIE1996, Moultaka2004}, 
%spectroscopic observations, 
%we can try to detect and track the evolution of starspots and other magnetic phenomena on HD 134319 in precision.
HD 134319's stellar parameters were greatly summarized in 
SIMBAD\footnote{\label{ft:SIMBAD}\url{http://cdsweb.u-strasbg.fr/}}
and ExoFOP\footnote{\label{ft:exofop}\url{https://exofop.ipac.caltech.edu/tess/target.php?id=202426247}}
and some of them are listed in table~\ref{tab:info}. 
%
%% Basis
% - Age
% Montes2001: Late-type members of young stellar kinematic groups. 
% McCarthy2001: rot-per & HK emission ~ 300 Myr old Ursa Majoris association, space motion ~ 625 Myr old Hyades cluster, intrisic X-ray luminosity ~ 200*L_sun. >> can younger than 300 Myr.
% Decin2003: HD 134319 is an element of the Hyades supercluster (600 Myr) (Montes et al. 2001)
The age of HD 134319 was firstly deduced as $0.625\,\mathrm{Gyr}$ by classifying it as a member of the Hyades supercluster in terms of space motion by \citet{Montes2001} and \citet{McCarthy2001}, while other values from high-resolution echelle spectra were determined by \citet{Valenti2005} ($6.3\,\mathrm{Gyr}$), \citet{Takeda2007} ($2.36\,\mathrm{Gyr}$) and lastly \citet{Isaacson2010} ($0.06\,\mathrm{Gyr}$).
% - GJ 577 B/C.
Besides, a young nearby companion binary system, with GAIA magnitude of $G = 14.25$, and separation of $5.39\pm0.02\,\arcsec$ from HD 134319, was detected by infrared imaging \citep{McCarthy2001, Mugrauer2004} and adaptive optics \citep{Lowrance2003} (see section \ref{sec:GJ577BC}). 
% - RV
The radial velocity (RV) of HD 314319 was firstly determined as $V_{\mathrm{R}}=-3.8\,\mathrm{km/s}$ by \citet{1950ApJ...111..221W}, and most recently derived from OHP/ELODIE data with value $V_{\mathrm{R}} = -6.362 \pm 0.011\,\mathrm{km/s}$, which was included in catalogue of RV standard stars by GAIA \citep{Soubiran2018}. Its long-term stability from 2002 to 2013 was reported by \citet{Butler2017} in their RV exoplanet survey using precision RV measurements by Keck/HIRES.
%
%Consequently, HD 134319 is likely to be a young single star from the spectroscopy, while its companion can contaminate the flux in wide-field photometry like TESS which will be discussed in section \ref{sec:GJ577BC}.

%% spec studies: emission, RV.
% Soderblom1985: Chromospheric Ca ii H & K emission (R_HK) and rotation (w.r.t. mass, age) @solar-type stars
% Duncan1991: S_HK = 0.415        # data table.
% Wright2004: S-index: B-V=0.677, Mv=5.17, S=0.412, std(S)=4.14%, logR_HK=-4.35, Prot=5day, log(Age/yr)=8.28
% Valenti2005: stellar prmt: Teff, logg, M/H, Na/H, Si/H, Ti/H, Fe/H, Ni/H, vsini, VR ...
% Takeda2007: table of stellar PRMT.
% Lopez-Santiago2010: spectroscopic data catalog: RV, vsini, space motion, EW-Caii HK, EW-Ca IRT, EW-Li, R_HK.
% Butler2017: 20 yr survey of RV by LCES @HIRES-Keck. S-index, H-index (Ha)	# series of value (16 measurements)
% Mittag2018: S-index, R_HK.  # statistical relation: activity-rotation-Rossby	# S=0.419(0.032), Per=4.43, B-V=0.677 from (D91) Duncan et al. (2005), (W04) Wright et al. (2004), (I10) Isaacson & Fischer (2010), (W07) White et al. (2007), (W11) Wright et al. (2011).
% Morris2019: connection between starspots and chromo-act @ARC 3.5m tele-Apache Point Obs (R~31500). # S=0.42(0.02), logR_HK'=-4.12(0.02), logg=4.59, vsin=10.9, Teff=5762(40)
The chromospheric activity of HD 134319 was recognized by several studies \citep{Soderblom1985, Duncan1991, Wright2004, Lopez-Santiago2010, Butler2017, Morris2019}, by measuring its Ca II H \& K emissions in terms of S-index and/or its corresponding derivative $\lg R_\mathrm{HK}^\prime$, giving typical median values of $S \sim 0.42$ and/or $\lg R_\mathrm{HK}^\prime\sim-4.12$, which indicates HD 134319 to be likely an ultra-active star \citep{BoroSaikia2018}.
%Together with HD 134319 and other numerous objects, \citet{Soderblom1985} and \citet{Mittag2018} investigated statistical relations among activity level, rotation, colour and age, and recently \citet{Morris2019} reported the positive connection between activity level and spot coverage by estimating TiO molecular band absorption. However, \citet{Morris2019} also noted HD 134319 as an outlier with "very small spot coverage" ($f_s = 0.00\pm0.06$) which significantly violates their conclusion.
While the spot coverage on HD 134319 was reported to be very small through estimation of TiO molecular band absorption by \citet{Morris2019} in their investigation of the connection between activity level and spot coverage, they noted HD 134319 as an outlier with high activity but small spot coverage.
%We also want to note the works of \citet{Wright2004} and \citet{Butler2017} who presented time series of S-index based on spectra included in our analysis.

%% phot studies.
% Messina1998b: Youth & high level of chrom & phot mag activity
% Messina1998: LCI, 2 long-lasting active longitudes, area >=16% surface @incl=90
% Messina2001: rotation-activity connection: Max V-band LC amp & period
% Morris2019: connection between starspots and chromo-act
% Alekseev2006: Review: methods(spec, phot, polariztion), active longitudes, DR, Cycle.
The starspots on HD 134319 from photometric analysis were firstly proposed by \citet{Messina1998b} who clarified the increase of peak-to-peak amplitude of LC towards decreasing wavelengths to the presence of of spots. Soon \citet{Messina1998} used rotational modulation to predict long-lasting active longitudes and spot covering fraction of at least $fs \geq 0.16$ for this BY Dra type variable by employing light curve inversion to multi-band \citep[u, v, b and y,][]{Stromgren1966} photometric data spanning from 1991 to 1995. Later \citet{Messina2001} inverstigated the rotation-activity connection from photometry, including HD 134319. 

In a word, the youth and high level of chromospheric and photospheric magnetic activity on HD 134319 make it a good proxy for the young Sun not far after its arrival at the zero age main sequence.
In this paper we present a detailed analysis of starspot configuration, its evolution and chromospheric activity on HD 134319, using high-precision LC newly observed by TESS and collected spectroscopic data. % observed by Keck/HIRES and OHP/ELODIE. %, to derive starspots and their evolution.
In section \ref{sec:obs}, we introduce both the photometric and spectroscopic observations.
We describe our approach for sliding spot modelling and the method for measuring relative equivalent widths from spectroscopic data in section~\ref{sec:model}.
We then give the results and respective discussions in section~\ref{sec:discussion}, and finally the conclusions are summarized in section \ref{sec:summary}.

%% file: 2-observation-reduction.tex
\section{Data and REDUCTION}\label{sec:obs}
%% ======
%\textcolor{red}{\sout{HD 134319 has magnitude of $V = 8.41^{\ref{ft:SIMBAD}}$, radius of $R = 0.94\,R_{\sun}$ \citep{GAIA:2018:DR2:VizieR} and a mass of $m = 1.01\,m_{\sun}$. Its stellar parameters were greatly summarized in SIMBAD$^{\ref{ft:SIMBAD}}$ and ExoFOP$^{\ref{ft:exofop}}$ and some of them are collected in table~\ref{tab:info}. It was classified as a G5 BY Dra type main sequence variable with a moderate rotation.} }
%\uppercase\expandafter{\romannumeral5}

\begin{table}
    \caption{Stellar parameters of HD 134319 in literatures.}
    \label{tab:info}
    \begin{tabular}{ll}
        \hline
        Parameter           & Value \\
        \hline
        RA, DEC           & 15h05m49.90423, +64\degr02'49.9415"$^{\ref{ft:SIMBAD}}$                  \\
        $V$ mag             & V=8.41$^{\ref{ft:SIMBAD}}$               \\
        Radius ($R_{\sun}$) & 0.94$^a$, 0.92984998$^b$, 0.92985$^{\ref{ft:exofop}}$                      \\
        Mass ($M_{\sun}$)   & 1.01$^{\ref{ft:exofop}}$                          \\
        $T_{\text{eff}}$ (K)& 5668.04$^{\ref{ft:exofop}}$ , 5635.87$^a$, 5662$^c$, 5636$^d$     \\
        Metallicity [Fe/H] & 0.03$^{\ref{ft:SIMBAD}, \ref{ft:exofop}}$                                                 \\
        log$g$               & 4.50556$^{\ref{ft:exofop}}$    \\
        $v~\text{sin}i$ (km/s)     & 10.6$^c$, 17.89$^e$, 11.39$\pm$0.06$^f$, 10.9$^d$                   \\
        \hline
        %\multicolumn{2}{l}{$^a$ \citep{GAIA:2018:DR2:VizieR}  }\\
        %\multicolumn{2}{l}{$^b$ \citep{Stassun2018}    	    }\\
        %\multicolumn{2}{l}{$^c$ \citep{Valenti2005}           }\\
        %\multicolumn{2}{l}{$^d$ \citep{Morris2019}            }\\
        %\multicolumn{2}{l}{$^e$ \citep{White2007}             }\\
        %\multicolumn{2}{l}{$^e$ \citep{Lopez-Santiago2010}    }\\
    \end{tabular}
    \\
    $^a$\citep{GAIA:2018:DR2:VizieR}, $^b$\citep{Stassun2018}, $^c$\citep{Valenti2005}, $^d$\citep{Morris2019}, $^e$\citep{White2007}, $^e$\citep{Lopez-Santiago2010}.
\end{table}

\subsection{Spectroscopy}
%% ======
The spectroscopic data for HD 134319 collected online consist of two parts.
One is the observations publicly accessible and available for download from Keck Observatory Archive\footnote{\url{https://koa.ipac.caltech.edu/}}, which were carried out during years 1999 - 2013 by the HIRES spectrometer on the Keck \UpperRoman{1}-10m telescope \citep[Keck/HIRES,][]{Vogt1994}. 
%Keck/HIRES is a grating cross-dispersed, echelle spectrograph capable of observing fainter objecting with high resolution of about $R=67000$.
The spectral resolution is about $R=67000$.
%On 18 August 2014, Keck/HIRES underwent an upgrade of its CCD from a Tektronix 2048 CCD to a mosaic of 3 MIT/Lincoln Labs $2048\times4096\,\mathrm{px}$ CCDs which improves the pixel size and other performances\footnote{\url{https://www2.keck.hawaii.edu/inst/hires/hires_data.pdf}} and extends to cover the \Ha~wavelength.
%
The other is the data from ELODIE Archive\footnote{\url{http://atlas.obs-hp.fr/elodie/}} which was observed by ELODIE spectrograph on the 1.93m telescope of Observatoire de Haute-Provence \citep[OHP/ELODIE,][]{ELODIE1996, Moultaka2004}.
%OHP/ELODIE was designed to measure stellar RV with high precision at a
The spectral resolution is $42000$. %ranging from $3906\,\mathring{A}$ to $6811\,\mathring{A}$ on a $1024 \times 1024\,\mathrm{px}$ CCD \citep{ELODIE1996}.
Totally 26 spectra of OHP/ELODIE observed during years 1995 - 1997 and 22 spectra of Keck/HIRES observed during years 1999 - 2013 were collected.

Reduction of the raw data from Keck/HIRES was carried out using standard tasks in the {\sc IRAF} package\footnote{IRAF is distributed by the National Optical Astronomy Observatories, which is operated by the Association of Universities for Research in Astronomy (AURA), Inc, under cooperative agreement with the National Science Foundation.}, 
which basically included flat-field division, scattered light correction and wavelength calibration. 
%The wavelength calibration was done using ThAr arc lamps, which were taken before or after the exposure of the target.
%We want to note that t
The bias and dark corrections were omitted due to their absence in some spectra, and that flat corrections were done using flats with wide orders (there existed another type of flats with narrow order in a fraction of observations).
%there exist one kind of flats in a fraction of spectra with narrower order than normal one which are common to all observations.
%and that only the flats %taken at the beginning of night and 
%with wider spread of the traces which are common to all observations were included in the process. 

OHP/ELODIE provides two types of products: the combined one dimensional spectrum ({\it s1D}) which was uniquely resampled in wavelength and given in instrumental relative flux covering $4000$ to $6800\,\mathring{A}$, and the two dimensional image ({\it s2D}) which contains the extracted and deblazed spectrum in 67 orders.	%, $1026\times67$

According to wavelength coverage, we chose the spectral portions around \CaII~ H \& K, \Hb~ and \Ha~ lines. For Keck/HIERS, \Ha~is available since 18 August 2004. For OHP/ELODIE {\it s2D} data, \Hb~ portion was selected from order 32 (\Hb~in order 31 was omitted due to bad pixels %within absorption core, which also contaminates {\it s1D}),
\Ha~portion was selected from order 64, and \CaII~H \& K (in orders 3 and 2, respectively) were omitted due to low signal to noise ratio (SNR).

\subsection{GJ 577 B/C}\label{sec:GJ577BC}
%% ======
%% GJ 577 B/C. IR
% McCarthy2001: by proper motion, distance by HST, seperation 5.41 arcsecond west, Imag=13.0, M5.
% Lowrance2003: Gl 577 companion maybe bianary.
% Mugrauer2004: Optical spec of GJ 577 A & GJ 577 B	# McCarthy2001, Lowrance2003
% Burgasser2005: info of companion GJ 577 BC
% Morris2019
A young nearby proper motion companion of HD 134319, with separation of $5.39\pm0.02\,\arcsec$ to the west and GAIA magnitude of $G\sim 14.25$, was detected by infrared imaging \citep{McCarthy2001, Mugrauer2004} and adaptive optics \citep{Lowrance2003}. 
\citet{Mugrauer2004} determined its spectral type as $M4.5$, mass as $0.16-0.20\,M_{\sun}$, age as $> 100$ Myr and found its strong \Ha~emission and deep TiO and VO molecular absorption bands.
\citet{Lowrance2003} further recognized it as a binary system (GJ 577 B/C) with separation $0.082\,\arcsec$, both of which lie on the stellar/substellar boundary with spectral type between $M5$\UpperRoman{5} and $M6$\UpperRoman{5}.
This scenario of binary system was also noted by \citet{Burgasser2005} and \citet{Martin2017} and confirmed by proper motion anomaly \citep{Kervella2019}.

The separation of $5.39\,\arcsec$ is apparently far enough in spectroscopy on Keck/HIRES\footnote{\url{https://www2.keck.hawaii.edu/inst/hires/slitres.html}} and OHP/ELODIE\footnote{\url{http://www.obs-hp.fr/www/guide/elodie/elodie.html}} and thus unlikely to contaminate the primary's spectra.
However the companion is near enough in TESS photometry (spatial resolution of $21\,\arcsec/\mathrm{px}$\footnote{\url{https://heasarc.gsfc.nasa.gov/docs/tess/observing-technical.html}}) to be crowded in the same pixel as the primary.
%
%despite the low spatial resolution of TESS ($21\,\arcsec/\mathrm{px}$\footnote{\href{https://heasarc.gsfc.nasa.gov/docs/tess/observing-technical.html}{https://heasarc.gsfc.nasa.gov/docs/tess/observing-technical.html}}), due to the following reasons:
%However it lies almost in the same pixel as primary on TESS CCD and contributes definitely to the measured flux.
%1. Considering that the distance between HD 134319 and the Sun is about $45.9\,\mathrm{pc}$ \citep{2018yCat.1345....0G}, a distance of $100\,\mathrm{AU}$, corresponding to an orbital period of more than $1000\,\mathrm{yr}$, of the companion from its primary, and a separation of about $4\,\mathrm{AU}$, corresponding a possible orbital period of about $20\,\mathrm{yr}$, between members within the companion system itself, can be estimated, according to solar system. The potential periodic signal with such a long timescale is hard to detect by TESS.
%2. The magnitude of companion was reported by GAIA with $G\sim 14.25$, which is fainter than the primary ($G\sim 8.23$) by two orders.
% makes it unlikely that it is a significant source of contaminating flux
%Considering that GJ 577 B/C has a possible period of $\sim20$ years (separation of $0.082\,\arcsec$ in distance $\sim 45.9\,\mathrm{pc}$ from the Earth), and that their
For simplicity we assume that it will not put detectable distortion, but at most a small offset about $0.40\%$, on resultant TESS LC because it is much fainter ($G = 14.25$) than the primary ($G = 8.23$) as measured by GAIA \citep{GAIA:2018:DR2}.

\subsection{TESS Light Curve}\label{sec:TESS_LC}
%% ======
%TESS observes the sky in sectors measuring 24◦ × 96◦ that extend from near the ecliptic equator to beyond the ecliptic pole. Each sector is observed for two orbits of the satellite around the Earth, or about 27 d. Sectors begin to overlap towards the ecliptic pole which means that at mid-latitudes the same star will be observed in more than one sector.
% observations of TIC 278956474 in two minute cadence data, processed by NASA’s TESS Science Processing Operations Center (SPOC) (Jenkins et al. 2016; Jenkins 2019).
% Balona2019: Light curves are generated with two-minute cadence using simple aperture photometry (SAP) and pre-search data conditioning (PDC). The PDC pipeline module uses singular value decomposition to identify and correct for time-correlated instrumental signatures in the LCs. In addition, PDC corrects the flux for each target to account for crowding from other stars and their effects. Only PDC light curves are used in this paper.
% TESS-PDCSAP, crowds @ different sectors, check by hand: nothing apparently different => no contanmation found. ==> use PDCSAP.
% The TESS 2-min light curves of the remaining sources were visually inspected by eye individually to determine those that showed any signs of rotational modulation.

The TESS mission \citep{Ricker2015} was designed to observe an extremely wide sky area in sectors measuring $24 \times 96\degr$, utilizing four wide-field cameras aligned in a mosaic, extending from near the ecliptic equator to beyond the ecliptic pole. 
The initial two-year observation strategy of TESS includes 26 sectors, each of which is observed in two highly elliptical $13.7\,\mathrm{day}$ orbits around the Earth, spanning about $27$ days.
TESS provides simple aperture photometry (SAP) and pre-search data conditioning (PDC) processed LC generated from two-minute cadence data, processed by NASA's TESS Science Processing Operations Center (SPOC) pipeline \citep{Jenkins2016} which is an successor of Kepler's pipeline\footnote{\url{https://heasarc.gsfc.nasa.gov/docs/tess/}}, as well as the calibrated target pixel file (TPF, which contains the pixel level photometric time series).
All of them are publicly accessible and available for download from Mikulski Archive for Space Telescopes (MAST) at STSci data base\footnote{\url{https://mast.stsci.edu/}} to ensure simultaneous usage of final products and check of the data quality in pixel level.

Considering that the PDC process was primarily designed for planet hunting but not for the study of stellar variabilities which usually show non-unique and much broader periodic durations and variations than planetary transits, and could be ignored or misinterpreted by the pipelines, one is generally recommanded to check the systematic correction by PDC and redo it when necessary \citep{pyke3_2, pyke3_1}.
We used calibrated TPFs to check whether the pipeline chosen apertures, which are different between sectors, can overcome the possible contamination from the crowed neighbours and temporal variation of photons distribution, i.e., whether the PDC processed LC is suitable for stellar activity investigation.

%The process was carried out using Python package {\sc Lightkurve}\footnote{\url{https://github.com/KeplerGO/lightkurve}} \citep{Lightkurve} which is launched in Jupyter Notebook environment\footnote{\url{https://github.com/jupyter/notebook/}}. {\sc Lightkurve} was built on libraries including {\sc NumPy}, {\sc SciPy} and {\sc Matplotlib} and relative to {\sc astropy} \citep{astropy:2018}, {\sc astroquery} \citep{astroquery:2019}, {\sc celerite} \citep{celerite:2017} and {\sc tesscut} \citep{tesscut:2019}.

%\subsubsection{Photometric apertures}\label{sec:aperture}
%% ====== obs-info, crowds, check-mask, resultant-LC.
The determination of optimal aperture in wide field-of-view photometry like TESS \citep{Ricker2015} is non-trivial due to the crowdedness of objects versus wide spread image of stars on CCD.
Crowdedness contributes to extra flux on target and should be avoided by employing smaller aperture, which, however, causes wastage of the target flux and leads to lower SNR.
Estimations of \textit{flux fraction} and \textit{crowding matrix} were done by PDC to recover the real brightness of target in Kepler and later TESS \citep{Batalha2010}.
However the automatically predetermined aperture and its correction in pipeline might be unsuitable for individual case, as widely revealed (for example, the large jumps between adjacent sectors) on many targets \citep[e.g.][]{Ozavci2018, Xu2021}, and should be checked with caution, especially in investigation of stellar variation.

HD 134319 was observed by TESS in sectors 14--16 and 21--23 during its second year's routine. 
Besides GJ 577 B/C, there are other three neighbours (1620024559829845504, 1620023597757174144 and 1620024692973605760) matched with \textit{Gaia} DR2 \citep{GAIA:2016, GAIA:2018:DR2} around HD 134319 recorded on the CCD image of TPFs.
The last two are far enough from HD 134319 and properly excluded by PDC. The first one ($\text{Gmag} = 13.9574$, fainter than HD 134319 by about two orders of magnitude) was excluded outside the PDC apertures in sectors 14--16, but not %vice versa
in sectors 21--23 due to the variation of flux spread.
Considering that no recognizable difference in profile of whole LC between with and without this neighbour was found, we concluded that it dose not contribute detectable contamination on HD 134319's LC.

Moreover, we assessed the validity of PDC determined apertures by comparing the LC profiles derived from different apertures. % both within each sector and between sectors.
One can manually define a smaller TPF masks, i.e. photometric apertures, by setting a larger value of parameter \textit{threshold}, which defines the threshold in choosing the pixels to be integrated in {\sc Lightcurve}, and vice versa. We found that the profile of integrated LC was stable with masks both slightly smaller and larger than the PDC, indicating that the spread function of HD 134319 was stable over time and the apertures determined by PDC were suitable for our study.

\begin{figure*}
	\includegraphics[width=0.95\textwidth]{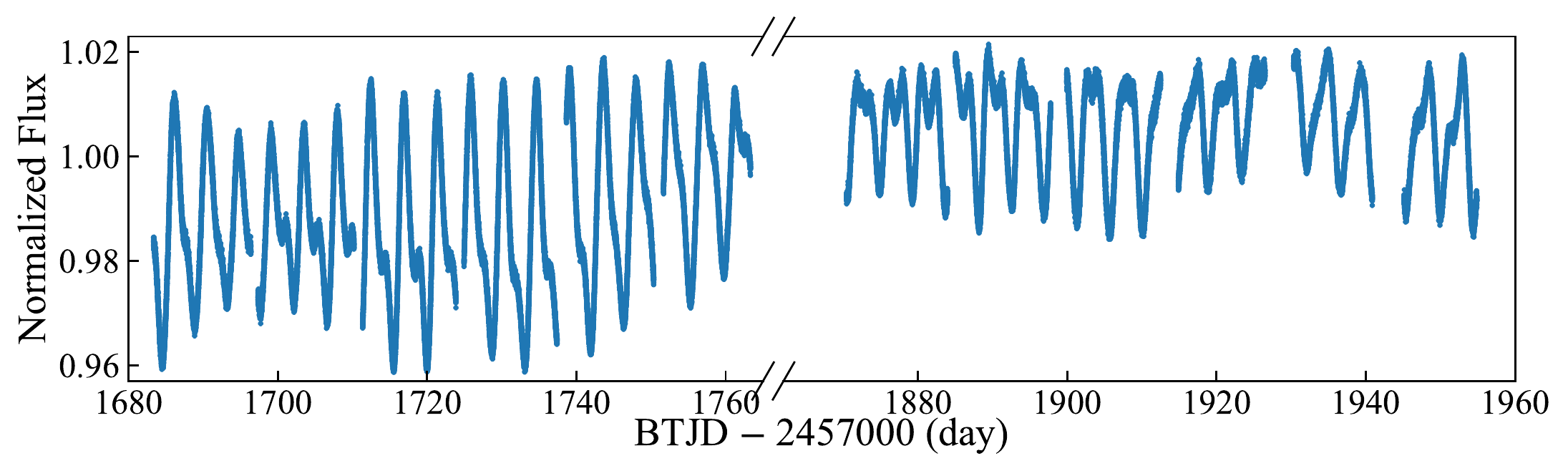}
	\caption{Resultant light curve.}
	\label{fig:lc}
\end{figure*}

Consequently, we used the PDC corrected LC, i.e. \textit{PDCSAP} long cadence data, in our analysis. There was no empirical evidence of improper correction by PDC for all sectors except two parts with improper offsets. One is the first half of sector 22 (1899 -- 1927 after $\mathrm{BTJD}-2457000$) and was subtracted by an offset of $400$ ADU, the other is sector 23 and was added by an offset of $1000$ ADU.
%\textcolor{red}{\sout{This kind of offset was employed to ensure that the profiles between adjoint LC segments coincide with each other, and was artificial to some extent due to too long rotation period versus time length of data to stitch sectors using semi-quantitatively approach \citep[e.g.][]{Ozavci2018}. However this does not make too much difference on starspot evolution because the offset attributes only to the stable deviation of amplitude of rotational modulation, which results in either the variation of spot size or additional polar spot.}}
This kind of offset, to remove the discontinuities between the improper segment and its surrounding LCs, was estimated qualitatively due to the too short time length of segment to adopt a semi-quantitative method \citep[e.g.][]{Ozavci2018}. The possible under- or over-estimation of the offset attributes to a common deviation of LC, which results in either spot size variation or additional polar spot but has not much to do with the spot evolution.
At last the LC was normalized to unity by dividing its median over the whole time span (figure~\ref{fig:lc}).

%% file: 3-model.tex
\section{Model and analysis}\label{sec:model}
%% ======

\subsection{Modelling the light curve}
%% ======
Periodic or quasi-periodic variation in photometric LC is generally attributed to the starspot rotating into and out of view from Earth \citep{Valio2017}. As a manifestation of internal magnetic flux on the stellar surface, starspot generates notable distortion on LC which makes it as good indicator of the internal magnetic structure and tracer in measuring the surface rotation. However, reconstructing two-dimensional distribution of the starspots from one-dimensional disk-integrated photometric time series is always not easy due to the weak constraints from the observation itself \citep{Lanza2016} as well as the large parameter space and high degeneracy between parameters of spot model.
%in the mathematical literature of spot modelling.
One should to reduce the number of parameters as small as possible in spot modelling and treat the results with caution.

\subsubsection{Rotational period of HD 134319}\label{sec:per}
% ======
%The rotation period of HD 134319 was measured by different authors. Table \ref{tab:per} provides some of those results and their reference. 
In literatures (table~\ref{tab:per}), the rotation period of HD 134319 was reported as $4.448\,\mathrm{day}$ by \citet{Messina1998b} and $4.43\,\mathrm{day}$ by \citet{Wright2011} from photometric observations, and estimated as $5.0\,\mathrm{day}$ by \citet{Wright2004} and $3\,\mathrm{day}$ by \citet{Isaacson2010} from statistical determination of chromospheric activity fluctuations.
With the favour of long-term precise photometry by TESS, we can obtain an accurate measurement of the period revealed by the rotational modulation. We used the generalized Lomb-Scargle periodogram \citep[GLS, ][]{Zechmeister2009} to measure the period of time series in sectors 14--16 and 21--22 of TESS (sector 23 was temporarily excluded due to its empirically questionable LC) as $P=4.436391\pm 0.00137\,\mathrm{day}$, which is slightly larger than the one of \citet{Wright2011} and smaller than the one of \citet{Messina1998}. %We pretend it to be a reasonable estimation for our analysis.

\begin{table}
    \caption{Rotational periods of HD 134319 measured by different authors.}
    \label{tab:per}
    \begin{tabular}{lll}
        \hline
        Reference               & Period (days)           & Memo                    \\
        \hline
        \citet{Messina1998}$^a$ & $4.448\pm0.005$        & $^{b}$			\\
        \citet{Wright2004}      & $5.0$                  & cf. $R_\mathrm{HK}^{\prime}$		\\
        \citet{Isaacson2010}    & $3$                    & cf. $R_\mathrm{HK}^{\prime}$		\\
        \citet{Wright2011}      & $4.43$                 & $^{c}$						\\
        This work               & $4.436391\pm0.00137$   & $^{d}$ 						\\
        \hline
        %\multicolumn{3}{l}{$^a$ Also in \citet{Messina1998b, Messina2001}.}		\\
        %\multicolumn{3}{l}{$^b$ photometric Scargle-Press period \citep{Scargle1982}.}		\\
        %\multicolumn{3}{l}{$^c$ Cited by \citet{Butler2017, Mittag2018, Morris2019}. }\\
        %\multicolumn{3}{l}{$^d$ photometric period by FEPS \citep{Meyer2006}. }\\
        %\multicolumn{3}{l}{$^e$ photometric period by GLS \citep{Zechmeister2009}. }\\
    \end{tabular}
    $^a$Also in \citet{Messina1998b, Messina2001},
    $^b$photometric Scargle-Press period \citep{Scargle1982},
    $^c$photometric period by FEPS \citep{Meyer2006}, this value was cited by \citet{Butler2017, Mittag2018, Morris2019},
    $^d$photometric period by GLS \citep{Zechmeister2009}.
\end{table}

\begin{figure}
	\includegraphics[width=\columnwidth]{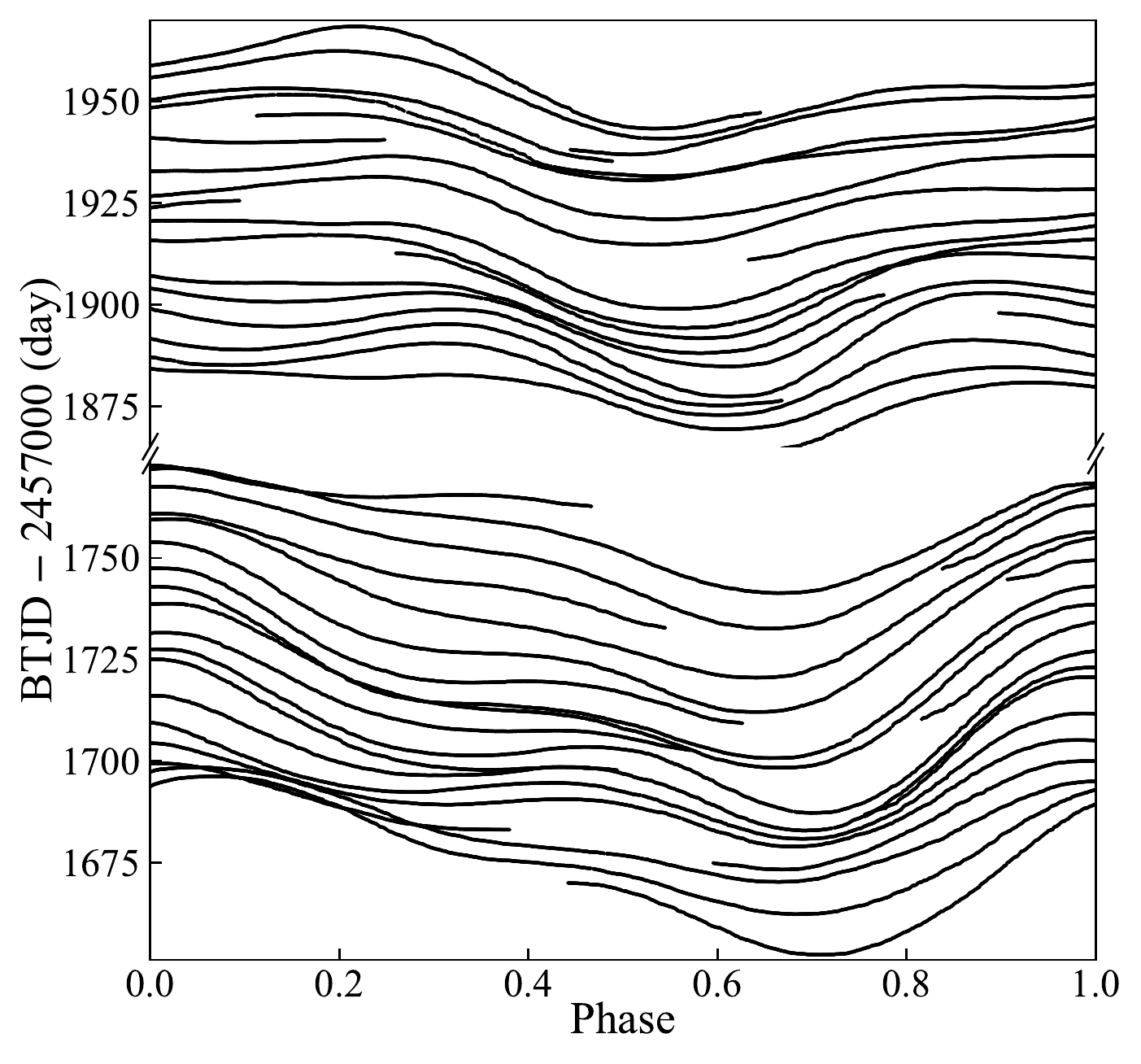}
    \caption{Phase-folded, median smoothed LCs for HD 134319 with GLS estimated rotation period $P = 4.436391\,\mathrm{day}$, showing a larger, primary "P" plus a smaller, secondary "S" condensed regions under starspot modulations and their notable evolution over time.
    The vertical position for each curve corresponds to the start time of chunk. Each time window is scaled to larger amplitude for easier view by subtracting $1$ then multiplying it by $800$, i.e. $f^{\prime} = 800(f-1)$.
    The long-time gap from $1763.5$ to $1869.5$ is marked by breaks in vertical axes.
    }
	\label{fig:lc_slice_vertical}
\end{figure}

In figure~\ref{fig:lc_slice_vertical} we show the phase-folded LC of HD 134319 using this rotation period.
%\textcolor{red}{\sout{It can be easily seen the spot configuration consisting of two condensed active longitudes.}}
Two dips in brightness per rotation with different depths are visibile throughout the observation age.
 The primary dip %\textcolor{red}{\sout{in brightness}} 
 ("P") is visible over the whole observation season, starting from $\text{Phase} \sim 0.7$ at onset and shifting gradually to $\text{Phase} \sim 0.5$ in the end. The secondary dip ("S"), emerging at $\text{Phase} \sim 0.3$ at onset, is recognizable before time $1763.5$, then becomes ambiguous after that due to its small amplitude. On the other hand, notable decrease in brightness loss of both "P" and "S" can be found with increasing time, revealing notable evolution of spot configuration.

%By the way, there is no photometric transit signals as predicted$^{\ref{ft:exofop}}$ can be recognized by eyes.

\subsubsection{\sc GEMC\_LCM}
% ======
As an analytical model towards photometric observation, light curve modelling \citep[LCM,][]{Budding1977,Dorren1987} simulates the LC with assumption that the drop in light intensity is caused by a small number of circular spots occupying on the stellar photosphere and thus is capable of reducing the parameter space to some extent.
% \citet{Budding1977} simulated the theoretical light intensity $I_c(t)$ (with time $t$) under modulation of non-overlapping spots as
%\begin{equation}
%    \label{eq:lcm:Ic}
%    I_c(t) = U \left[1-\sum_j^{N_{\text{spots}}}{(1-\kappa_{w_j}) {3\over{3-u}} [(1-u)\sigma_{0,j}^0+u\sigma_{1,j}^0]}\right]
%\end{equation}
%Where $U$ is the unspotted intensity, $u$ represents the linear limb-darkening effect, $\kappa_{w_j}$ is the spot-to-photosphere intensity ratio corresponding with the temperature contrast between $j$-th spot and the stellar photosphere, then the combined LC was obtained by summing over spot. The fundamental formula $\sigma_{m,j}^n$, called "$\sigma$-integrals", was defined by an integration over covered area, of the $j$-th spot, projected on the line of sight.
%\begin{equation}
%    \label{eq:lcm:sigm_mn}
%    \sigma_m^n = 1/\pi \int\int_{\text{spot\ area}}{x^m z^n dxdy}
%\end{equation}
%Here the projection was done in $xyz$ Cartesian coordinate system orients the z-axis towards the observer, and can be translated to the spherical polar system of spot longitude $\lambda$ and latitude $\beta$, corresponding with given inclination of stellar rotation axis $i$ and spot radius $\gamma$, as $\sigma_m^n \equiv \sigma_m^n(i, \lambda, \beta, \gamma)$. Specially $\sigma_0^0$ represents the projected area of the spot coverage, while $\sigma_1^0$ represents the effect of linear limb darkening.
%
LCM consists of parameters relative to the star (the unspotted intensity $U$, the linear limb-darkening effect $u$ and the inclination $i$) and individual circular spot (the spot-to-photosphere intensity ratio $\kappa_{w}$, the spot latitude $\beta$, longitude $\lambda$, radius $\gamma$ and its rotational modulation period $P$).
Practically stellar parameters $i$, $U$ and $u$ are constants and parameter $\kappa_w$ can be approximately constants and common to all spots.
%And the geometry on each spot can be described by three parameters, i.e. $\lambda$, $\beta$ and $\gamma$.
Taking into account the rotation, the periodical variation of longitude can be better defined as a function of of time $t$ by the rotational period $P$ and initial longitude $\lambda(0)$ as $\lambda(t) = 2\pi t/P-\lambda(0)$ for convenience \citep{Xu2021}. %Hence the number of independent parameters for each spot is $4$.

An efficient program, called "{\sc GEMC\_LCM}" \citep{Xu2021}, which was designed to inherit both the superior global optimization power of genetic algorithm and the high efficiency on parameter space exploration of Markov Chain Monte Carlo algorithm, was employed for spot modelling using LCM of \citet{Budding1977}.
{\sc GEMC\_LCM} is able to run parallelly on mulit-core CPU with the favour of {\sc OpenMP}.

\subsubsection{Parameter degeneracies}
% ======
Besides the large parameter space, the parameter degeneracies in spot modelling are another kind of problem in photometry modeling, which block the algorithm converging to the true solution. % during iteration. 
Practically it is difficult to overcome this problem in spot modelling in the absence of independent constraints.
However it is reasonable to divide such degeneracies into two types according to their sources, i.e., they are inherent or produced. The former comes from the model itself, while the latter comes from measurement related effects which can be suppressed by high precision photometry.

An inherent degeneracy between stellar inclination and spot latitudes was identified from numerical simulation by \citet{Walkowicz2013}, who also showed the possibility of failure in detecting the differential rotation when LC is dominated by one large active region.
Actually, when it comes to the case of LCM with single spot, one can easily find that this degeneracy can be expressed analytically as
\begin{equation}
    \label{eq:lcm:degeneracy}
    %I_c(t; i, U, u; \kappa_w, \lambda, \beta, \gamma) = I_c(t; 90\degr-|\beta|, U, u; \kappa_w, \lambda, \mathrm{Sign}\left[\beta\right](90\degr-i), \gamma)
    I_c(i, \beta) = I_c\left(90\degr-|\beta|, \mathrm{Sign}\left[\beta\right](90\degr-i)\right)
\end{equation}
Where $I_c$ is the simulated LC, $\mathrm{Sign}\left[\beta\right]$ means the sign of $\beta$, and specially $\mathrm{Sign}\left[0\right]=\pm1$.
In other words, the spot at latitude $\beta$ with inclination $i$ creates definitely the same LC as spot at latitude $\mathrm{Sign}\left[\beta\right](90\degr-i)$ with inclination $90\degr-|\beta|$.

%A produced degeneracy between spot configurations with different latitudes due to the low constraint of photometry on latitude versus noise and contaminations was noted as an "ill-posed" problem \citep{Lanza2016}. 
A produced degeneracy between spot latitude and radius due to the low constraint of photometry versus noise and contamination was noted as an "ill-posed" issue \citep{Lanza2016}.
% discrepancy , for example between spot configurations with different latitudes,
%Practically, distinction between LC with spots at different latitudes is usually subtle and thus likely to be submerged in noise or contamination.
%Measurements suffer from the observational accuracy and noise which block the algorithm from discrimination of subtle but important distinction between nearby solutions. 
%For example, simulations by \citet{Walkowicz2013} noted the increment of difficulty in distinguishing LCs between different stellar inclinations with increasing photometric noise, and \citet{Ioannidis2016} revealed that one can not constrain spot latitude, even with precise observation like Kepler 210, in the case of high ($\ge 70\degr$) stellar inclination.
Fox example, a larger spot at higher latitude distorts the LC with similar amplitude as a smaller spot at lower latitude by means of projection to light of sight.
However, it is possible to minimize such a defect to some extent with the mercy of high precision photometry as they lead to subtle but different profiles in principle \citep{Walkowicz2013}.
Besides, the contamination coming from the residual systematic errors in long-term observations such as Kepler and TESS can mixed with stellar variations in multiple timescales. Thus it is usually recommanded to do the systematic error corrections, circumstances alter cases, in stellar variation investigations \citep{pyke3_2, pyke3_1}. %We chose the scheme in preparing LC as described in section~\ref{sec:aperture}.
We have checked the LC as described in section~\ref{sec:TESS_LC}.

To derive the properties of spot configuration independent on parameter degeneracies, it is feasible to employ some assumptions a priori.
A predefined stellar inclination is capable of breaking its degeneracy with spot latitude.
The stellar inclination can be calculated from measured $v\sin i$, stellar radius $R$ and rotation period $P$ by
\begin{equation}
    \label{eq:i}
    i = \arcsin\left( v\sin i*P / {2 \pi R} \right)
\end{equation}
With the measured period $P=4.436391\,\mathrm{day}$ as derived in section~\ref{sec:per}, the stellar radius in table~\ref{tab:info} implies a maximal $v\sin i \sim 10.7\,\mathrm{km/s}$. Considering that the projected rotational velocity $v\sin i$ was estimated in the range between $10.6\,\mathrm{km/s}$ and $17.89\,\mathrm{km/s}$ (table~\ref{tab:info}), HD 134319 is preferred to have a high inclination.
As a comparison, we adopted fixed inclination of $i=75\degr$ and $i=90\degr$ in modelling.

\subsubsection{Surface differential rotation}
% ======
The differential rotation was thought to play a critical role in generating and maintaining the stellar magnetic field. 
The surface differential rotation (SDR) on the Sun was observed in relative motion of sunspots and can be expressed by a quadratic law
\begin{equation}
    \label{eq:SDR}
    P(\beta)=P_{\mathrm{eq}}/(1-\alpha \sin^2 \beta )	
\end{equation}
Where $P(\beta)$ is the stellar rotation period at latitude $\beta$, $P_\mathrm{eq}$ is the reference period at equator, and $\alpha$ is relative rate representing the strength of SDR, which was measured as $\alpha=0.2$ on the Sun. A similar law of SDR is generally assumed on stars other the Sun by analogy \citep{Henry1995}. By this definition a zero $\alpha$ means a rigid rotation, a positive $\alpha$ represents a Solar-like SDR while a negative $\alpha$ represents an anti-Solar SDR.

\subsubsection{Spot modelling}
% ======
%The existence of multiple periods in photometric time series was widely used as the signature of differential rotation 
% ===
%To quantitatively trace the starspot evolution and SDR on HD 134319, we should determine the precise sizes and positions of the starspots over time. 
The whole LC was split into chunks with $5$ days duration each, slightly longer than one rotation period, and advanced by $1$ day forward, providing a total of $129$ such chunks for spot modelling. In such a short timescale, the spot positions are not expected to evolve, while their radii are allowed to change linearly indicated by the apparent variation of LC within adjoint rotations. %Each chunk was advanced by $1\,\mathrm{day}$, providing a total of $129$ such chunks in 6 sectors of observation for spot modelling.

The modelling was done under hypothesis of a stellar sphere with uniform surface brightness occupied by circular dark spots with uniform temperature.
%The other probable effects are omitted, noting as argued by \citet{Ioannidis2020} that inclusion of bright regions (or plages) in model does not lead to less extreme values for the spot sizes and temperatures.
The pre-determined parameters were chosen as follows.
The limb darkening effect was assumed to be a linear law with coefficient %fixed to an approximation value of 
$u=0.5119$ from the table of \citet{Claret2018} for a star with $T_{\mathrm{eff}}\sim5600\,K$ and $\log{\it{g}}\sim4.5$. The uniform spot-to-photosphere intensity ratio was fixed to $\kappa_w=0.22$ empirically derived from statistics (figure 7 of \citet{Berdyugina2005}). The stellar surface brightness was also fixed as the maximal value of LC, $U\sim1.02$, which is artificial to some extent because the target is always occupied by spots. At last, by comparison, we employed two values of stellar inclination $i=75\degr$ and $i=90\degr$.

The spot configuration in each chunk was derived under a two-spot model.
The number of spots was determined by two reasons. One is that there are two minima in almost all chunks, probably inferring a distribution of two concentrated active regions on the surface. The other comes from our tests in checking the global optimizing capability of models with two and three spots. It was found that three 
(or more)
spot model provided slightly better fiting as expected, while resulted more likely in families of solutions with equally good fits which reduces its reliability, comparing with two-spot model.
Therefore a two-spot model was sufficient to obtain acceptable fits.
Note that, strictly speaking, the real number of spots (or active regions) might or might not eqaul to the number of minima in LC \citep{Jeffers2009, Basri2020}, but we would rather to fit the LC with a simple spot-model for efficiency and convergence of optimization, and if the spot distributions on the star can be described by such a two-spot model it would be the result.

The determination of rotational modulation period corresponding to spot is non-trivial. 
%The existence of SDR on stars can be detected by measurements of the rotational periods of spots at different latitudes, which can be derived directly by LCM in terms of periodically varying longitude over time. 
Usually it can be determined with high precision due to the strong constraint of long-term photometric time series on longitude, despite exceptions, e.g. when one spot dominates the LC \citep{Walkowicz2013}.
%However, situation changes when we do spot modelling on short-term timescale over single chunk as the constraint on period becomes weak. 
However, such a constraint becomes weak in our case when we do spot modelling for single chunk, i.e. in pretty short timescale.
Instead we adopted fixed period in modelling after several numerical experiments, and thus the derived longitude might vary due to two effects, i.e. the evolution of spot longitude, and the difference between the chosen and actual rotational modulation period of spot.

As a conclusion, we employed a two-spot model to inverse the longitudes, latitudes, linearly evolving radii of spots (i.e. eight free parameters in total) for each chunk, using fixed period $P=4.436391\,\mathrm{day}$.
For the sake of higher probability of globally optimized solution, each chunk was fitted for many runs, and the results were then divided into groups corresponding to different spot configurations. % by requiring that adjacent chunks have similar solutions as they overlap by $\~80\%$ of data points.
Finally we elected the group with minimal combined residuals $\sum_j(\chi^2_j)$ summing over all chunks ($j$) as the final result.

\subsection{Relative variations of chromospheric activity}\label{sec:model_ew}
%% ======
Core emissions in strong optical spectral lines, such as the \CaII~H \& K and \Ha~lines, are proven proxies for magnetic flux on the Sun \citep{Eberhard1913}. Their strengths as well as variations were used for monitoring stellar activity levels and detecting long-term activity cycles similar to the solar $11\,\mathrm{year}$'s cycle.
For example, the S-index, developed by the Mount Wilson Observatory HK Project \citep{Duncan1991}, is a measure of the emission in the \CaII~ H \& K line cores %from singly-ionized calcium 
in the lower to middle chromosphere due to magnetic heating, and was widely used in measuring the stellar rotation period as well as the long-term activity cycles \citep[e.g.][]{Hempelmann2016, Butler2017, Mittag2018, Mittag2019}. % Butler2017

\subsubsection{Approach}\label{sec:ew_approach}
% ======
According to the wavelength coverage of spectra, we adopted the \CaII~H \& K, \Hb~and \Ha~lines to measure chromospheric activity. The spectral subtraction technique \citep{Barden1985, Montes1995} has been widely used in measuring their emissions with the favour of comparison star or synthesized spectrum \citep[e.g.][]{Montes1995, Gu2002, Cao2019}.
However, noting the difficulty in finding the continuum around the \CaII~H \& K lines \citep{Shkolnik2005} while the requirement of precise measurement on temporal variations of chromospheric activity indicators in our case, we alternatively employed a scheme based on \citet{Xu2021}, which is capable of deriving relative variation of equivalent widths of chromospheric activity indicators in series of spectra with high precision. % by subtracting an overall spectrum from all spectra.

% === modified scheme.
The scheme consists of following steps. 
$\mathbf{1}$. Delete outliers in each spectrum contaminated by cosmic rays, telluric lines, CCD bad pixels, etc.	% , residual instrumental defect
$\mathbf{2}$. Normalize the spectrum with highest SNR to unity as reference spectrum $F_R$. %Normalizations for spectrum around \Ha~and \Hb~can be done by continuum division with reliability. For \CaII~H and K lines, however, normalization was done respectively by fitting a straight line to the edges of spectral portion, $7\,\r{A}$ wide, centred on emission core \citep{Shkolnik2005}.
$\mathbf{3}$. Fit other individual spectra $F^{\prime}(\lambda^{\prime})$ by a product of $F_R(\lambda)$ and a polynomial $P(\mathbf{a}; \lambda-\lambda_0$), i.e. $F^{\prime}(\lambda) = P(\mathbf{a}; \lambda-\lambda_0)F_R(\lambda)$ where $\lambda_0$ is the reference wavelength, and then apply a shift $\Delta\lambda$ due to RV, i.e. $\lambda^{\prime} = \lambda+\Delta\lambda$.
% Here the parameters to be determined are $\mathbf{a}$, the parameter array in polynomial fitting, and RV shift $\Delta\lambda$. 
Each individual spectrum is inversely transformed to unity $F_U$ using derived parameters $\mathbf{a}$ and $\Delta\lambda$.
%Such a process is capable of reducing the adverse influence of possible large noise in $F^{\prime}$ which results in unreasonable deviation of $P$ if we otherwise directly fit $F^{\prime}$ to $F_R$.
$\mathbf{4}$. The overall spectrum $F_M$ is obtained as the median over a subset of spectra with high SNR. Then the residual spectrum is $f = F_U - F_M$. 
$\mathbf{5}$. Correspondingly the relative equivalent width is $\Delta W = W - W_M = -\int f d\lambda$. By fitting $f$ with a Gaussian function, $g(\lambda) = A \exp\left[-(\lambda-\lambda_0)^2/(2\sigma^2)\right]$, we have $\Delta W = -\sqrt{2\pi}A\sigma$.
Note that minus value means emission.

\begin{figure}
  	\includegraphics[width=\columnwidth]{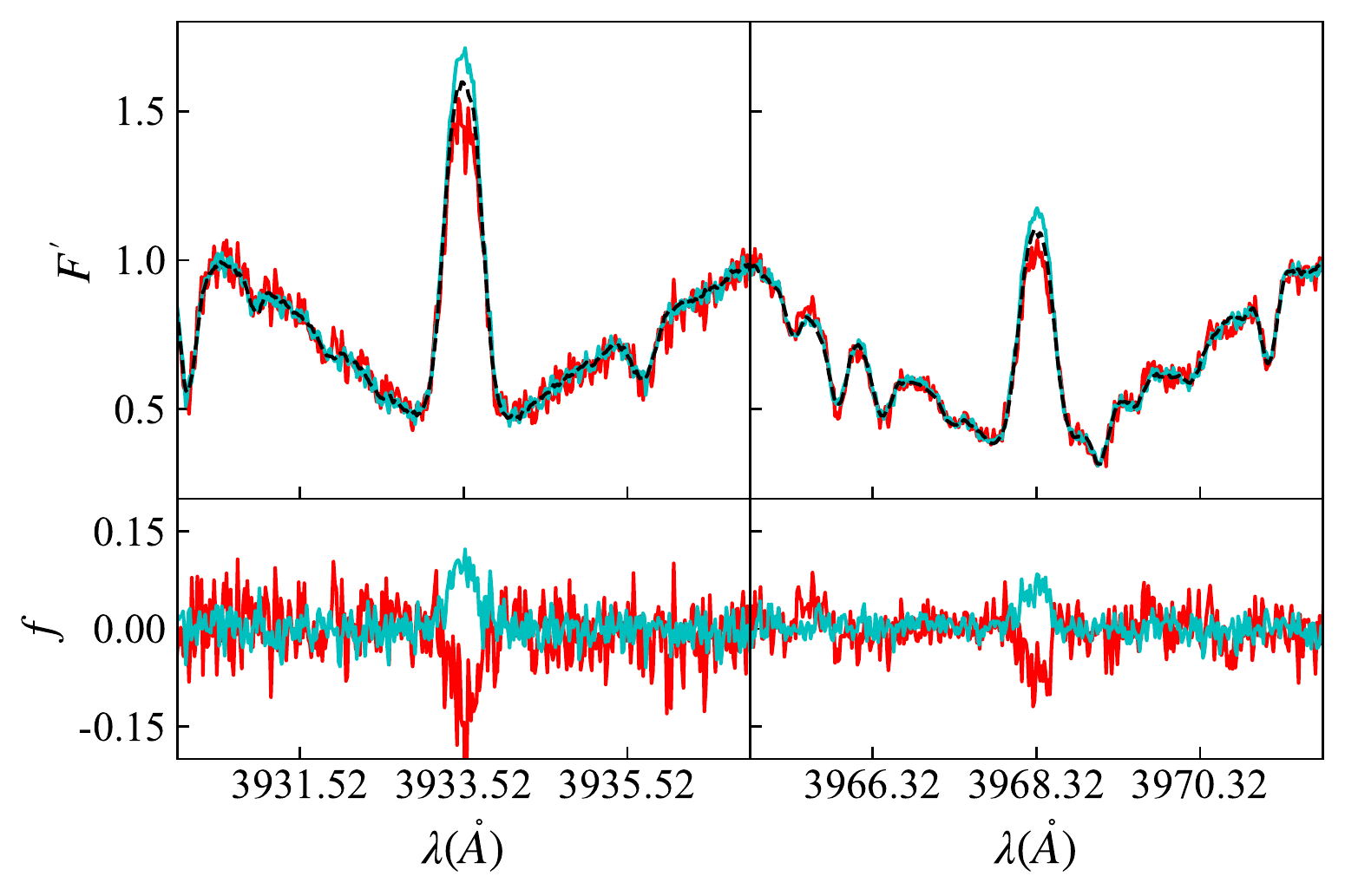}
  	\includegraphics[width=\columnwidth]{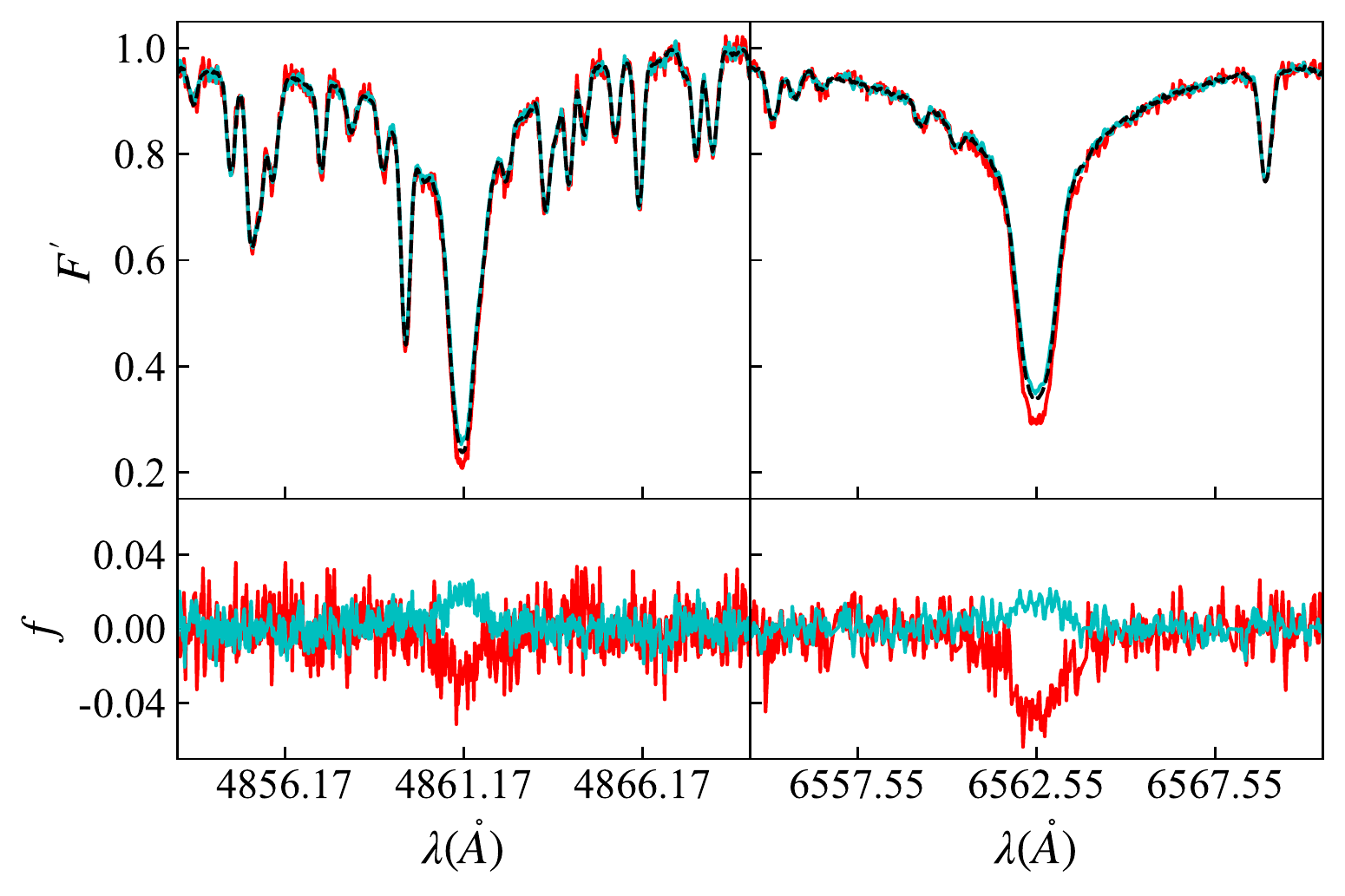}
  	\caption{Examples of spectra observed at 20060416 (red line) and 20110615 (cyan line) reduced by our scheme. Each panel shows spectra $F^{\prime}$ derived by step $\mathbf{3}$ with the overall spectrum $F_M$ overlaid by black dashed line, along with their respective residual spectra $f = F^{\prime} - F_M$ derived by step $\mathbf{5}$.}
	\label{fig:ew_scheme_eg}
\end{figure}

Figure~\ref{fig:ew_scheme_eg} shows two example spectra observed by HIRES/Keck at 20060416 and 20110615 (table~\ref{tab:ew}), when all of \CaII~K, H, \Hb~and \Ha~emisson lines exhibited almost the most significant negative and positive residuals, respectively. The relative variation was small compared to the overall spectra but can be extracted with reliability by above scheme.

\begin{figure}
	\includegraphics[width=\columnwidth]{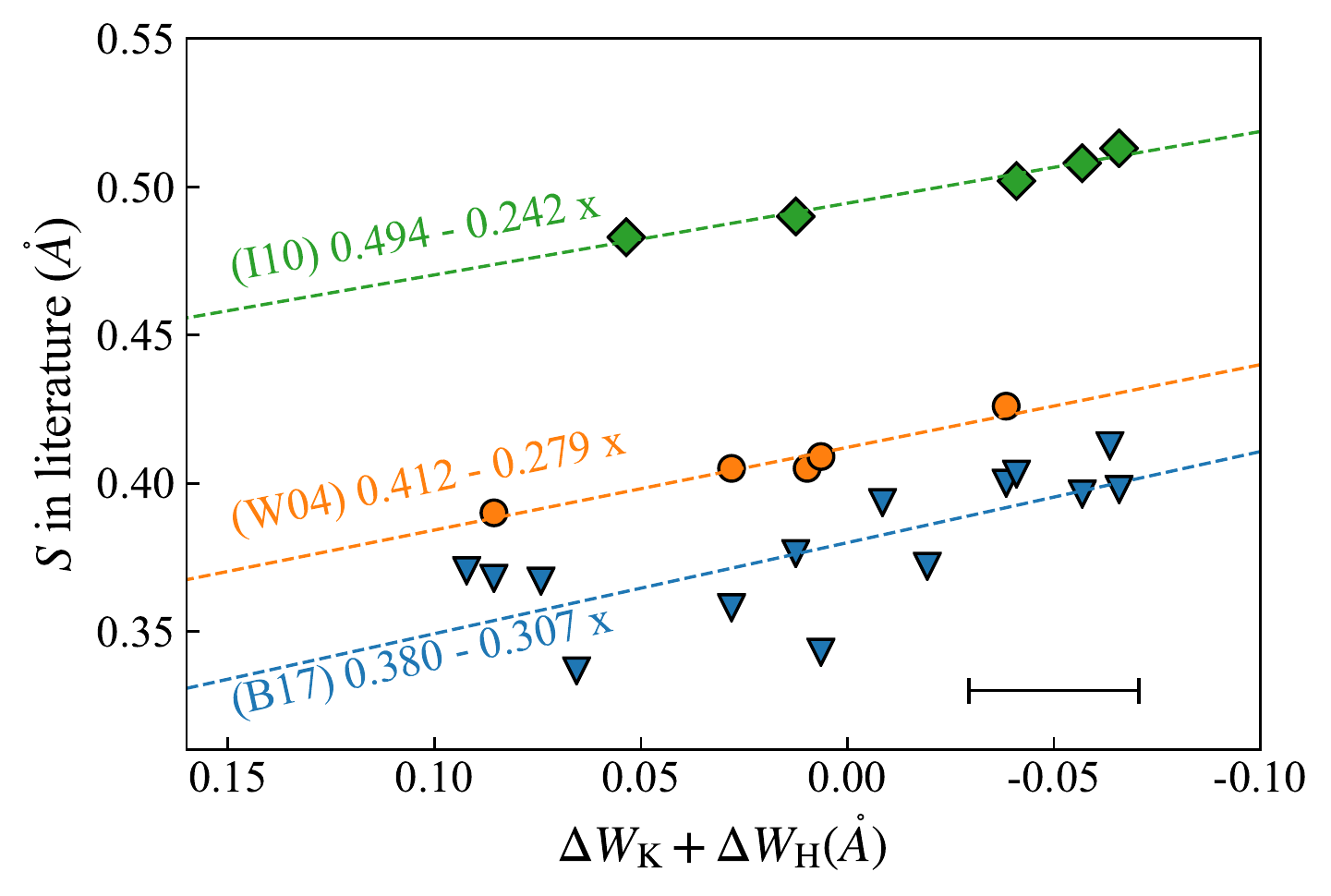}
   \caption{Comparisons between our measurements, $\Delta W_{\text{K}} + \Delta W_{\text{H}}$, and previous measurements of S-index from \citet{Wright2004} (W04), \citet{Isaacson2010} (I10) and \citet{Butler2017} (B17). Linear fittings to the three cases are overlaid. The horizontal error bar represents the typical uncertainty of our measurement.
    }
	\label{fig:ew_comp}
\end{figure}

By using subsets of Keck/HIRES spectra, \citet{Wright2004}, \citet{Isaacson2010} and \citet{Butler2017} measured S-index of HD 134319 independently. Comparison between our measurements and their results is shown in figure~\ref{fig:ew_comp}, overlaid with respective linear fittings. Note that \citet{Wright2004} employed a scheme similar to us by using the highest SNR observation as a template in measuring the "sensitive differential S-values". Our result fits well with the ones of \citet{Wright2004} and \citet{Isaacson2010}, indicating the reliability of our scheme in measuring the relative variation of chromospheric activity indicators.

\subsubsection{Residual spectra}
% ======
The parameters in above scheme were determined by numerical experiments and listed in table~\ref{tab:ew_portions}.
In step $\mathbf{2}$, The reference spectrum was the one observed at Jan 27, 2008 (record 41 in table~\ref{tab:ew}) by Keck/HIRES.
In step $\mathbf{3}$, the background portions, whose length should be short enough for a good fitting by a low-order polynomial while long enough to include enough absorption lines to measure RV shift, were chosen as a compromise and the gap centred on respective emission core was temporally included for background fitting.
In step $\mathbf{4}$, the subset used in calculating the overall spectrum $F_M$ consists of records 31--37, 39--41, 43, and 46--48 (table~\ref{tab:ew}) with high SNRs and common to all indicators. 
In step $\mathbf{5}$, the width of Gaussian function, in accordance with $\sigma$, was estimated by plotting residual spectra $f$ together and fixed as $\sigma = \mathrm{FWHM}/2.35482$ for each index. %, where $\mathrm{FWHM}$ means full width at half maxima, 
The fitting interval within which Gaussian fitting was done was used to estimate the fitting errors.

\begin{table*}
    \caption{Spectral fitting parameters.}
    \label{tab:ew_portions}
    \begin{tabular}{lcccccc}
        \hline
        Index	& $\lambda_0$		& \multicolumn{2}{c}{Parameters in step $\mathbf{3}$ (Polynomial fit)}	& \multicolumn{2}{c}{Parameters in step $\mathbf{5}$ (Gaussian fit)} \\
                & ($\mathring{A}$)	& Background portions ($-\lambda_0$)	& Order	& Fitting interval ($-\lambda_0$) & $\sigma$ \\ 
        \hline
        \CaII~K	& $3933.52$ & $-7 $ -- $-0.75$ and $0.75$ -- $7 $ 	& $2$  & $ -7$ -- $7 $ & $0.140$ \\
        \CaII~H	& $3968.32$ & $-7 $ -- $-0.75$ and $0.75$ -- $7 $ 	& $2$  & $ -7$ -- $7 $ & $0.140$ \\
        \Hb~		& $4861.17$ & $-8 $ -- $-2   $ and $2   $ -- $8 $ 	& $3$  & $ -8$ -- $8 $ & $0.425$ \\			% 1/2.35482, 0.424660
        \Ha~		& $6562.55$ & $-13$ -- $-3   $ and $3   $ -- $13$ 	& $3$  & $-13$ -- $13$ & $0.828$ \\			% (6564.5-6562.55)/2.35482
        \hline
    \end{tabular}
\end{table*}

\begin{figure}
	\includegraphics[width=\columnwidth]{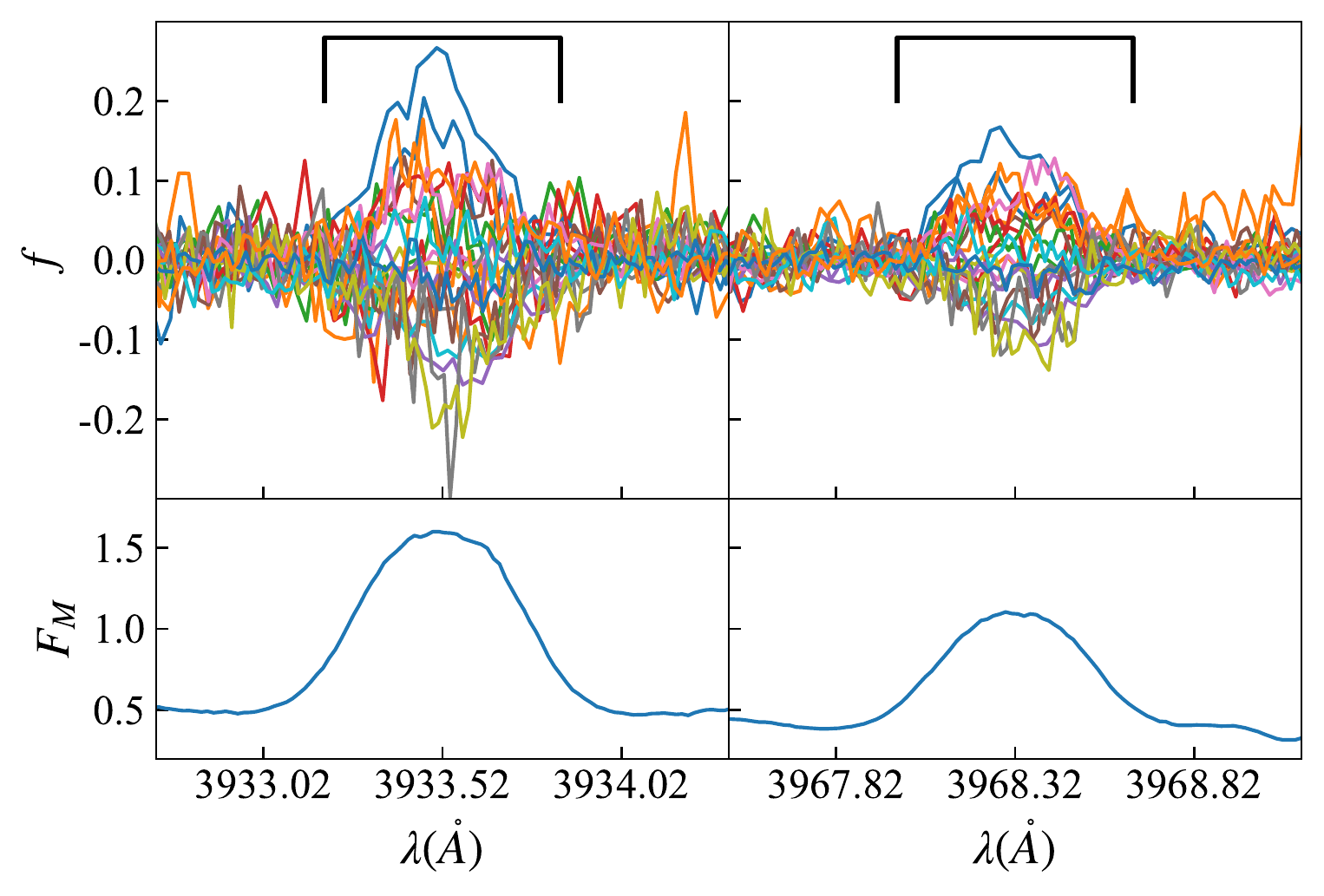}
	\caption{The residual spectra $f$ (up panel) and the overall spectrum $F_M$ (lower panel) ($f = F_U - F_M$) for \CaII~K (left) and H (right). The typical width of Gaussian fit (corresponding to $\sigma$) in measuring relative equivalent width $\Delta W$ was over-plotted by black lines.}
	\label{fig:ew_residual_CaHK}
\end{figure}

Normalizations of \CaII~H and K lines to unity through continuum division are usually questionable due to the difficulty in finding their continua, an alternative way was employed by fitting a straight line to the edges of spectral portion \citep{Shkolnik2005}, $7\,\mathring{A}$ wide, centred on emission cores, respectively.
Figure~\ref{fig:ew_residual_CaHK} shows the residual spectra $f$ and their respective overall spectra $F_M$ from Keck/HIRES.
Emissions of resonance from $F_M$ and variations of chromospheric activity from $f$ are notable, indicating a high level of magnetic activity.

\begin{figure}
	\includegraphics[width=\columnwidth]{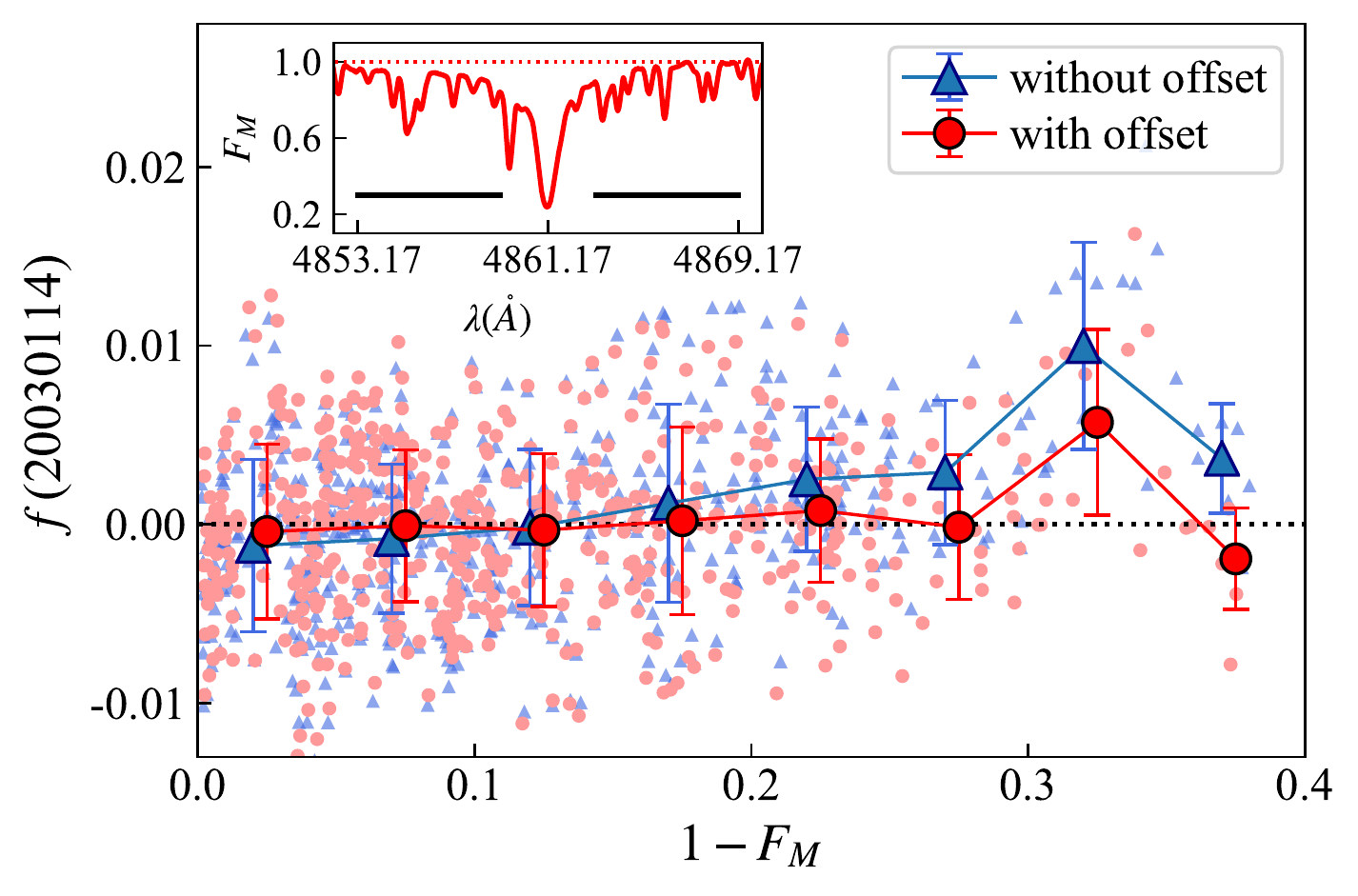}
	\caption{Comparison between residuals $f$ with and without an offset $o$ included in fitting $F^{\prime}(\lambda^{\prime})$ by $P(\mathbf{a}; \lambda-\lambda_0)F_R(\lambda)+o$ versus $1-F_M$: the example of \Hb~was observed by Keck/HIRES on Jan 14, 2003. The overlaid larger triangles and circles, horizontally misaligned from each other, were sliding average of respective smaller symbols, the error bar represents the standard deviation of corresponding interval.
    The overall spectrum is plotted in sub-window with background portions in table~\ref{tab:ew_portions} overlaid with black lines.
    }
	\label{fig:offset}
\end{figure}

\Ha~ and \Hb~ can be normalized to unity by general continuum division with reliability.
%Usually \Ha~and \Hb~are easier for SNR than \CaII~H \& K for later stars which are intrinsically brighter at redder wavelengths.
Moreover, with the high quality of \Hb~and \Ha~lines by Keck/HIRES, we recognized an improper trend between fitted residuals and spectral absorption depth, i.e. the background portions in residual $f$, which might result from improper reduction, e.g. scattered light residuals, reflected by high SNR data. 
An offset is capable of fitting out such a trend mathematically, i.e. $F^\prime(\lambda) \equiv o + P(\mathbf{a}; \lambda-\lambda_0)F_R(\lambda)$, and was included in processing \Hb~and \Ha~of Keck/HIRES spectra. Figure~\ref{fig:offset} shows example of \Hb~with the most notable difference between with and without offset observed at Jan 14, 2003.

\begin{figure}
	\includegraphics[width=\columnwidth]{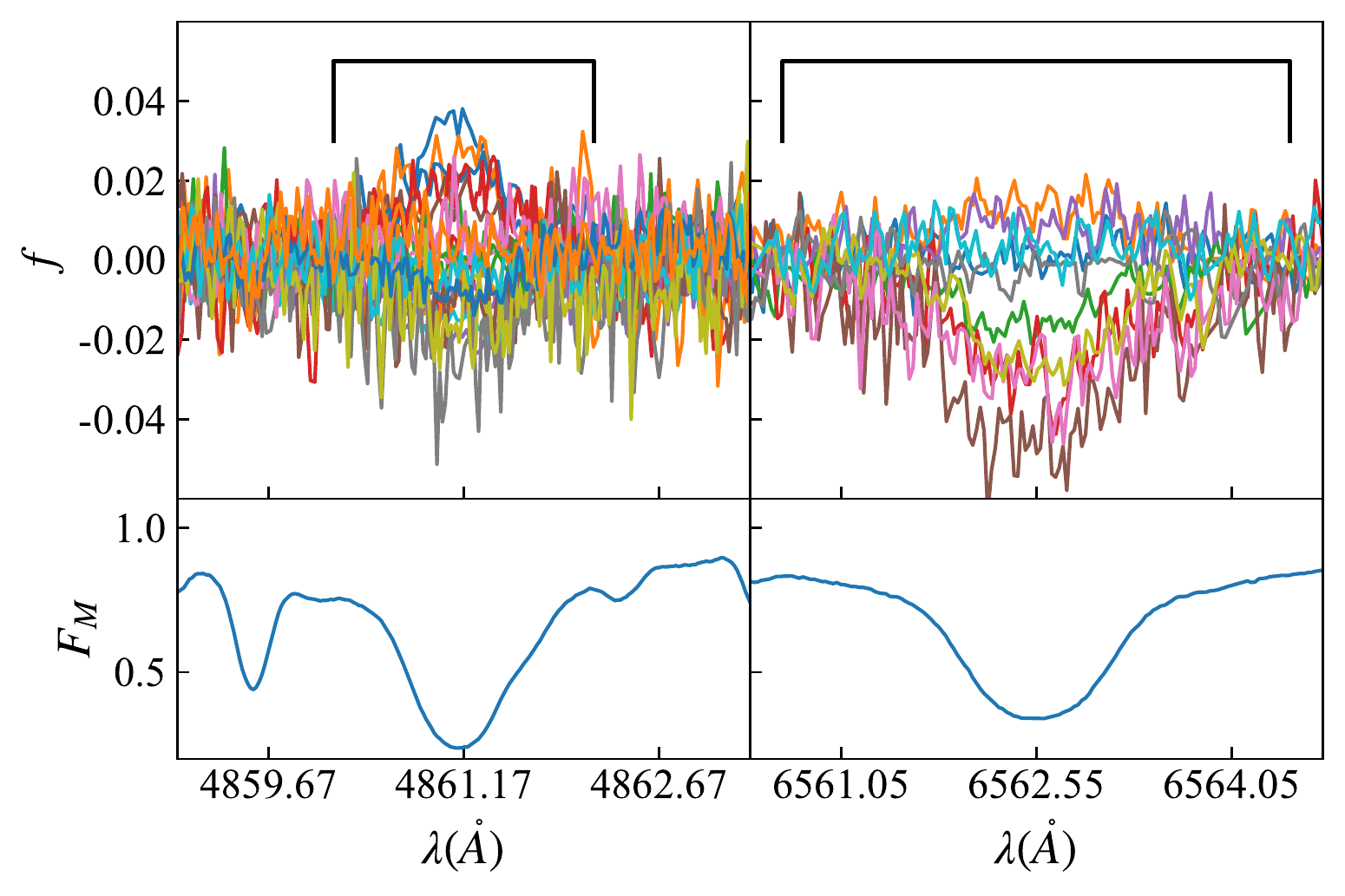}
    \caption{Same as figure~\ref{fig:ew_residual_CaHK}, but for \Hb~(left) and \Ha~(right).}
	\label{fig:ew_residual_Ha_Hb}
\end{figure}

Figure~\ref{fig:ew_residual_Ha_Hb} shows the residual spectra $f$ and respective overall spectrum $F_M$ of \Hb~and \Ha~lines, from which we can also see variable emissions with different amplitudes.

%% file: 4-result-discussion.tex
\section{Results and Discussion}\label{sec:discussion}
It's unfeasible to decide the real number of starspots from disk integrated photometric observation. The number of dips in LC corresponds only to the lower limit of starspots number, because LC can always be fitted by model with as many starspots as one likes \citep{Jeffers2009, Basri2020}. One has to adopt a priori compromise to reduce the free parameter space and increase the possibility of convergence in optimization so as to obtain a reasonable  solution in fitting. For example, LCM simulates the LC by an analytically model with a few circle spots, while light curve inversion (LCI) usually employs either the maximum entropy or Tikhonov criterion for a unique and stable solution (e.g. \citet{Lanza2006}). Thus, if the photosphere of HD134319 is covered by or equivalent to two dominant starspots (or active regions) versus surroundings with non-spot or stable distributed starspots contributing no rotational modulation, our result should reveal the distribution and relative evolution of starspots, or else it would rather be the description of hemispheric asymmetry.

%% ======
\subsection{Results comparison between different inclinations}\label{sec:rslt_comp}

Using the best-fit and manually checked solutions from many independent runs for each chunk, we were able to increase the probability of optimization to fall into global minima in limited iterations, and thus to trace the sizes and locations of starspots over the entire span of TESS data with reliability.

Practically, each chunk was modelled independently by $50$ runs for inclination $i=75\degr$ and $40$ runs for inclination $i=90\degr$.
The solutions could be divided into a few groups characterized by special spot configuration. %, all but one of which belong to local minima.
In most cases it was easy to elect the globally optimal solution by choosing minimal residual $\chi^2$, however outliers existed where local minima had equally good or even better fit than the predicted one. To overcome this problem we additionally assumed that spot configuration among adjacent chunks should be similar to each other as they shared about $80\%$ data points.

\begin{figure}
	\includegraphics[width=\columnwidth]{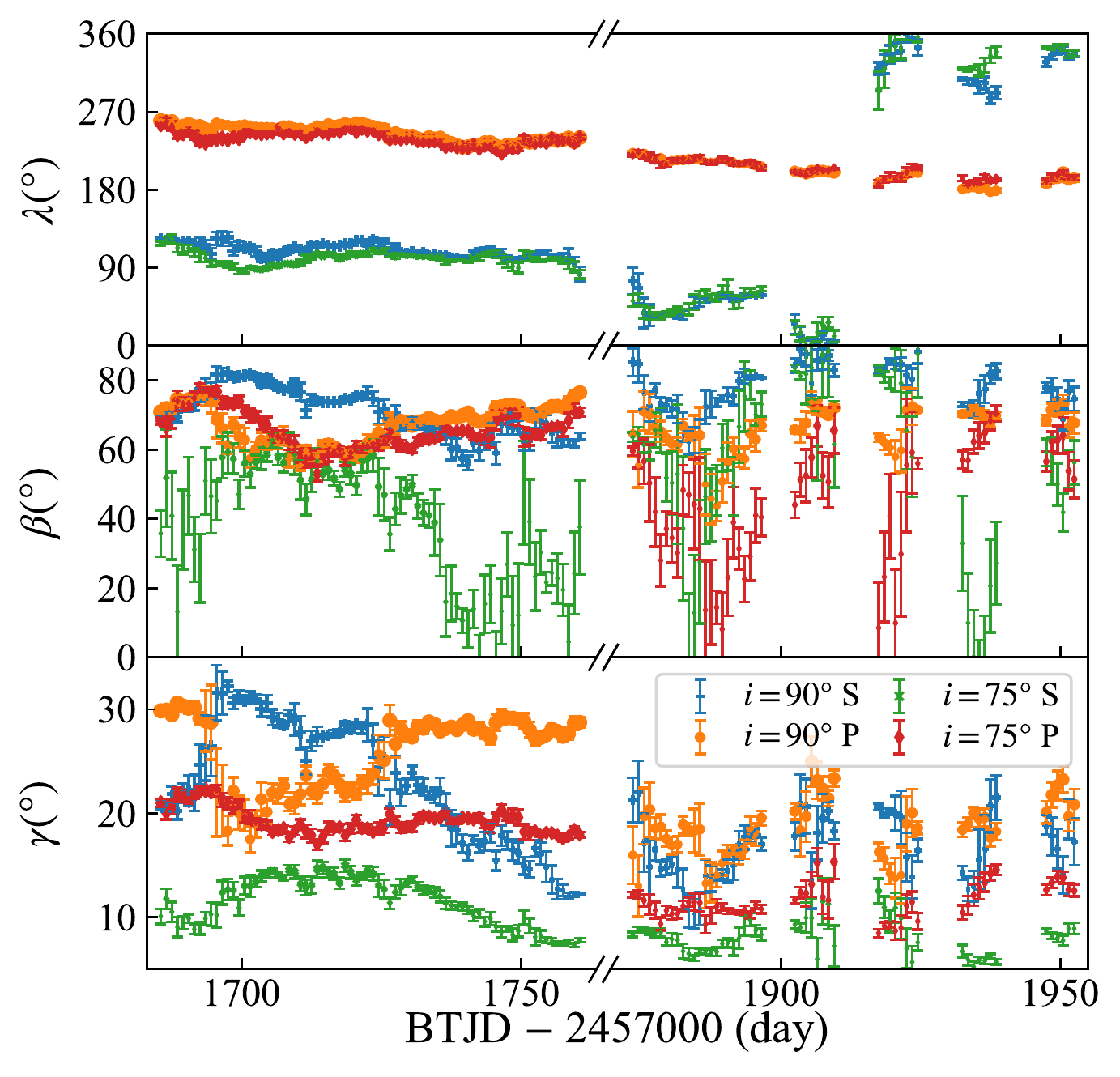}
    \caption{Temporal variations of spot longitudes (top), latitudes (middle) and radii (bottom) for inclination $i=75\degr$ (green crosses and red diamonds) and $i=90\degr$ (blue pluses and orange circles), respectively. The radius of each spot plotted, hereafter, takes its value at mid-epoch of respective chunk under the assumption of linearly evolving size. "P" represents the larger spot (primary) and "S" represents the smaller spot (secondary) in each case. The sizes of symbols are proportional to the spot radii.}
	\label{fig:LCM_comp}
\end{figure}

By comparison, figure~\ref{fig:LCM_comp} displays the spot longitudes, latitudes and radii corresponding with $i=75\degr$ and $i=90\degr$ respectively. Under assumption of linear evolution, we took the value of spot radius at mid-epoch of each chunk in plot and discussion hereafter.

The spot longitudes were derived with high stability for both cases as expected due to both the strong constraint of photometry on rotational phase and the orthogonality between stellar inclination and longitudes.
Two active longitudes can be easily recognized i.e. the primary one (named "P") centring around longitude $\lambda\sim 220\degr$ and the secondary one (named "S") varying around longitude $\lambda \sim 130 - -50\degr$.

Two features can be found from the distribution of spot latitudes.
First, the case of inclination $i=75\degr$ generally shows spots at lower latitudes than $i=90\degr$, which implies a mathematical deviation of configuration corresponding to input inclination, i.e., an evidence of parameter degeneracy.
Second, the result shows similar relative variations of spot latitudes between input inclinations, which indicates a possible constraint of high accuracy by TESS photometry on spot latitudes versus noise, to some extent.

\begin{figure}
	\includegraphics[width=\columnwidth]{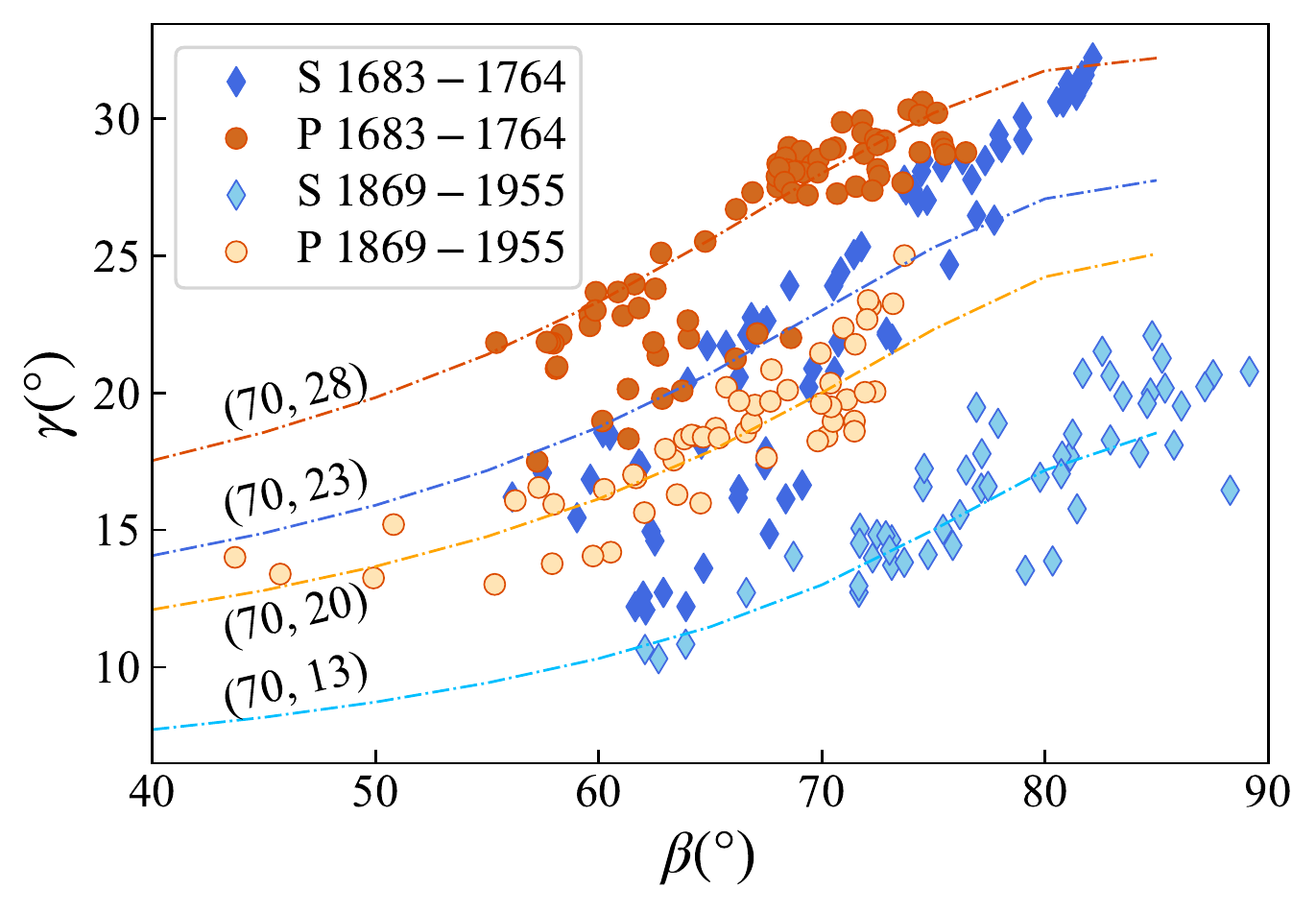}
    \caption{Spot radius--latitude ($\gamma-\beta$) correlation from best-fit solutions, different symbols represent spots "P" and "S" at different time ranges.
    The simulations for an artificial spot at latitude $\beta=70\degr$ initially with series radii $\gamma=13,20,23~\text{and}~ 28\degr$ through varying its latitude value from $\beta=40\degr$ to $85\degr$ are overlaid with dash-dot lines.
    %The coincidence between best-fit solutions and simulations indicates possible degeneracy between spot radius and latitude, while the deviation indicates the spot evolution.
    }
	\label{fig:LCM_rad_lat}
\end{figure}

A degeneracy between spot radius and latitude is recognizable in both cases.
Figure~\ref{fig:LCM_rad_lat} shows the distribution of spot radius versus its latitude for "P" and "S" in different time ranges for inclination $i=90\degr$, and their comparison to simulations for a supposed spot at latitude $\beta=70\degr$ initially with series radii $\gamma=13, 20, 23$ and $28\degr$ through setting the spot's latitude ranging from $40$ to $85\degr$.
The highly coincidence between results and simulations within time range 1869--1995 indicates probable existence of degeneracy, while the apparent deviation within time range 1683--1764 from such a coincidence means reliable variation of either spot radius or latitude.

\begin{figure}
	\includegraphics[width=\columnwidth]{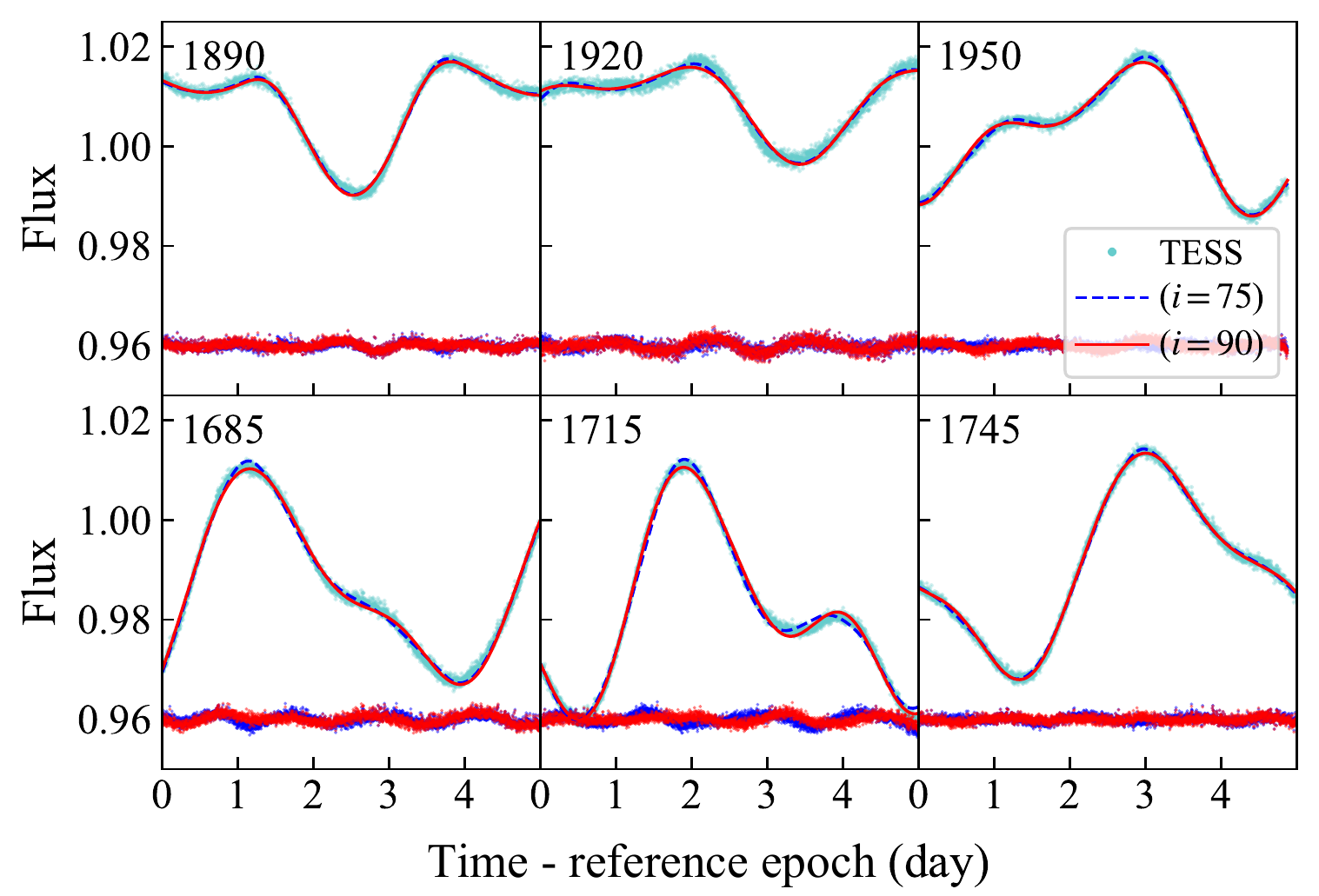}
    \caption{Illustration of modeling on inclinations $i=75\degr$ (blue dashed line) and $i=90\degr$ (red line) for TESS LC (cyan dots), one chunk for each sector, with epoch of each chunk indicated. The horizontal axis of each panel was modified to $0-5$ by subtracting its epoch. Residuals are over-plotted at value of $0.96$.}
	\label{fig:LCM_eg}
\end{figure}

Figure~\ref{fig:LCM_eg} shows one fitting example in each sector for cases of inclination $i=75\degr$ and $i=90\degr$.
Both cases reveal that good fits have typical residuals around $0.0001 - 0.0002$, implying that the spot configuration on HD 134319 can be reasonably described by two condensed active regions.
As a whole, fitting of $i=75\degr$ resulted in slightly larger $\chi^2$ than $i=90\degr$ in most chunks, which indicates an extremely high inclination of HD 134319, coinciding with the prediction from spectroscopy.
Hereafter spot modelling result with inclination $i=90\degr$ is employed in discussion.

%% ======
\subsection{Starspot Evolution and Differential Rotation}\label{sec:spot_evolve_DR}
% ===
%
% distribution:	1. two longitudes, high latitudes, large+small spots, 2. effect on GLS, 3. wrt dynamo.
% evolution:	1. longitudinal oscillation (avoid duplicate groups), wrt our knowledge , 2. decrease of radii, wrt activity level. 3. individual behavior (wrt spot lifetime)
% DR:			1. traditional scheme, 2. possible problem (longitude migration, evolution of radii, large over small, problem of model in deriving DR with only a few spots - ref2020), 3. Period, still left results.
%

\subsubsection{Configuration of two active regions}
\begin{figure*}
	\includegraphics[width=0.95\textwidth]{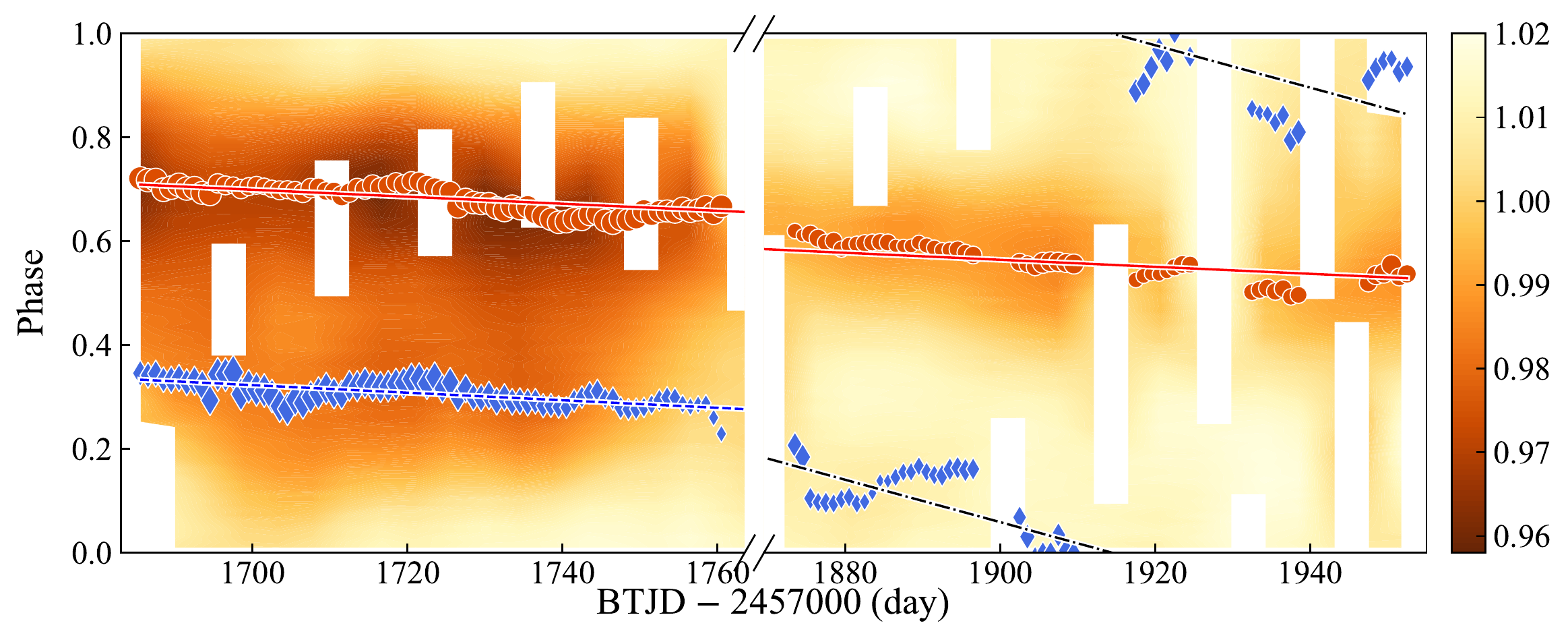}
    \caption{Continuous phased LC map of the entire TESS long cadence data set with period $P=4.436391\,\mathrm{day}$, with the spot longitudes from our two spot model as well as their linear fittings overlaid.
    The pixel shade, from dark to light, indicates the flux in each (time/phase) chunk, and is interpolated and plotted as contour map. Vertical white gaps correspond to times with no TESS data, and the long-time gap of observation are marked as breaks in horizontal axes.
    The longitude corresponding to the primary active region, "P", and its linear fitting are overlaid with orange circles and red line, while the longitude corresponding to the secondary active region, "S", and its linear fittings to the first and last half data separately, are overlaid with blue diamonds, blue dashed and dark dash-dot lines, respectively. The symbols are scaled proportional to the spot radii.
    }
	\label{fig:lc_slice}
\end{figure*}

In figure~\ref{fig:lc_slice} we reproduce the continuous phase versus time evolution map of flux, with spot longitudes from the best-fit solution and their linear fittings overlaid.
%We can manually identify two active regions that reveal gradually evolutions and characterize them by their amplitude of dip in brightness, i.e. the larger and primary one ("P") and the smaller and secondary one ("S"). 
%Under assumption of a high inclination of $i=90\degr$, these latitude
Two active regions characterized by their amplitudes of dip in brightness, i.e. the primary "P" located at phase about $0.71$ and the secondary "S" located at phase of about $0.34$, separating by $140\degr$ in longitude,
%
%From the best-fit solutions, the spot configuration on surface of HD 134319 during TESS era consists of two active regions centred on phase of about $0.34$ and $0.71$ at onset, i.e. separated by $140\degr$ on longitude.
are revealed to be located at high latitudes varying between $50$ and $80$ degrees (figure~\ref{fig:LCM_comp}). %, this is qualitatively acceptable for such a star with rapid rotation. %, however maybe questionable due to the weak constraint of photometry on latitude.
The solutions also indicate the difference between spot sizes, and the smaller one evolves more rapidly than the bigger one, despite the possible degeneracy between latitude and radius (figure~\ref{fig:LCM_rad_lat}).

% ref
In literature, the photometric study of HD 134319 was presented by \citet{Messina1998} and \citet{Messina1998b}, who analysed HD 134319 LC from 1991 to 1995 and reported a two spot configuration existed over $5$ years, despite the large gap between sections. 
% two spot configuration on stars:
% Davenport2015,2020: GJ 1243 (P0.592596d, m=0.24, near fully-convective): alpha^2-dynamo, very weak DR (alpha=0.00114, effect of DR), lat38.4 & primary & stable + lat0 & small & evolve spot, no cycle found.
% FloresSoriano2017: LQ Hydrae (P1.6d, dK2, BY Dra, ZAMS), chrom (Ca I 6439, Ca I 6718, Ha, Li 6708) spatial correlates with spots. # 3-spots by DI, only one dip in LC.
% Roettenbacher2013: KIC 5110407 (BY Dra, P3.4693, m=1.7, LCI, Kepler:100-180d), DR, no cycle found.
% Lanza2019: Kepler-17 (G2V, age<1.8Gyr, m=1.16, R=1.05, P=12.01, T=5780, i=89.88): 2 lon (one survive > 1400d), minimum DR, cycle ~ 48, 400-600d.
Such a configuration was widely found on other stars with different rotation periods.
Rapid rotator AB Dor ($P \sim 0.51479\,\mathrm{day}$, \text{K}0\UpperRoman{5}, $70\sim 100$ Myrs) was reported to maintain two long-lasting longitudes over tens of years from long-term photometry \citep[e.g.][]{Berdyugina2005AN, Ioannidis2020} and evolving SDR from Dopple imaging \citep[e.g.][]{Donati1997, Petit2002, CollierCameron2002, Jeffers2007} as well as ultra-active phenomena like superflares \citep{Schmitt2019}.
The rapidly rotating M4 type star GJ 1243 ($P \sim 0.59\,\mathrm{day}$, mass $\sim 0.24\,M_{\sun}$) was found to be dominated by one stable and another evolving active longitudes during Kepler and TESS epoch \citep{Davenport2020}.
Detailed studies on FK Com ($P \sim 2.4\,\mathrm{day}$, G4\UpperRoman{3}) revealed two active regions with apparently different sizes and their "flip-flop", phase-jumps, spot emerges, drifts and so on \citep[e.g.][]{Korhonen2001, Olah2006, Hackman2013}.
On BY Dra type star LQ Hydrae ($P\sim 1.6\,\mathrm{day}$, $\text{age} \sim 60\,\mathrm{Myr}$, K1\UpperRoman{5}), long-term photometry revealed dominance of two active longitudes over 20 years so that "the first evidence of flip-flop on single dwarf"  and multiple activity cycles, in both non-axisymmetric and axisymmetric (as in the Sun) modes, in time-length of years were reported \citep{Berdyugina2002, Lehtinen2012}.
%Spectroscopic studies revealed connection between photospheric spots and chromospheric active region \citep{FloresSoriano2017} and long-lived spots on both high-latitude and equatorial spots and that it was the high-latitude spot contributing to the photometric variations on LQ Hydrae \citep{ColeKodikara2019}.
Another analogy, Kepler-17 (with slower rotation, $P \sim 12.01\,\mathrm{day}$, G2\UpperRoman{5}) also showed two active longitudes, one of them survived over Kepler's observation period, and an evidence of cycle \citep{Lanza2019}.
Thus one can reasonably expect that such a two spot configuration exists widely on stars, and we can suggest an internal non-axisymmetric geometry for HD 134319, possessed by almost constant mean field configuration which was believed to indicate a small differential rotation \citep{Messina1998b}.

\subsubsection{Rotational modulation periods of starpots}\label{sec:spot_rot}
% evolution:	1. longitudinal oscillation (avoid duplicate groups), wrt our knowledge , 2. decrease of radii, wrt activity level. 3. individual behavior (wrt spot lifetime)
As shown in figures~\ref{fig:LCM_comp} and \ref{fig:lc_slice}, the spot configuration on HD 134319 revealed long-lasting features as well as significant evolution in their locations and sizes. For easier discussion, we divided the observation time into two durations, i.e. "T1", from 1683 to 1764 (sectors 14--16) and the other "T2", from 1869 to 1955 (sectors 21--23).
%The spot distribution on the global map (figure~\ref{fig:lc_slice}) suggests both long-lasting features and significant evolution in location and size of the most active feature.
%The long-lasting feature, the "P", appears to be stable and visible over whole observation era.
%On the contrary, the region "S" reveals rapid evolution in both longitude and size, especially during time 1689 to $1955$.
%According to the difference of spots evolution between different epochs, we divide the observation time into two durations, i.e. "T1", from 1683 to $1764$ and the other "T2", from 1689 to $1955$, for easier discussion.
%
The primary region "P" tended to be a long-lasting feature undergoing slow evolution from "T1" to "T2", % and thus could be properly described by a linear function,
while the secondary region "S" migrated not too much in "T1", like "P", but exhibit a much rapid evolution in "T2". %and thus it is reasonable to fit "S" by two linear fittings, 
%i.e. "S1" and "S2". %, corresponding to its evolution stages.

The rotational modulation period of individual spot can be measured from the phased LC \citep{Davenport2015} by fitting its longitude with a linear function
\begin{equation}
    \label{eq:P_spot}
    P_i = P_0/(1-m_i P_0)
\end{equation}
where $P_0$ is the phase folding period, the subscript $i$ represents the index of spot and $m_i$ is the slope. % and $P_i$ is the respectively resulting rotational period.
Note by this definition %the slope can be estimated by linear fitting of the spot longitude as overlaid in figure~\ref{fig:lc_slice} and 
a negative slope yields a smaller value than $P_0$.

The fittings are overlaid in figure~\ref{fig:lc_slice} with three lines.
Fitting of "P" gave %slope of $m_P = -0.0006748\pm0.0000171$, corresponding to 
an average rotational modulation period of $P_\mathrm{P} = 4.4236654\pm0.0003350\,\mathrm{day}$, starting at phase about $\lambda_\mathrm{P} = 0.71$ at epoch 1683 (red line).
Fitting of "S" in "T1" gave %slope of $m_\mathrm{S1} = -0.0007275\pm0.0000850$, corresponding to 
$P_\mathrm{S1} = 4.4226344\pm0.0016632\,\mathrm{day}$ and $\lambda_\mathrm{S1} = 0.34$ at epoch 1683 (blue dashed line).
And fitting of "S" in "T2" gave %slope of $m_\mathrm{S2} = -0.0040790\pm0.0003399$, corresponding to 
$P_\mathrm{S2} = 4.3580376\pm0.0064554\,\mathrm{day}$ and $\lambda_\mathrm{S2} = 0.19$ at epoch 1689 (dark dash-dot line).
$P_\mathrm{P}$ and $P_\mathrm{S1}$ were estimated with high confidence due to their stabilities, however $P_\mathrm{S2}$ was determined with larger uncertainty due to its rapid evolution.
Note that these values are smaller than the phase folding period, i.e. $P_0=4.436391\,\mathrm{day}$, derived by GLS determination of the whole LC. 

\subsubsection{Starspot evolution}\label{sec:spot_evolve}
\begin{figure}
	\includegraphics[width=\columnwidth]{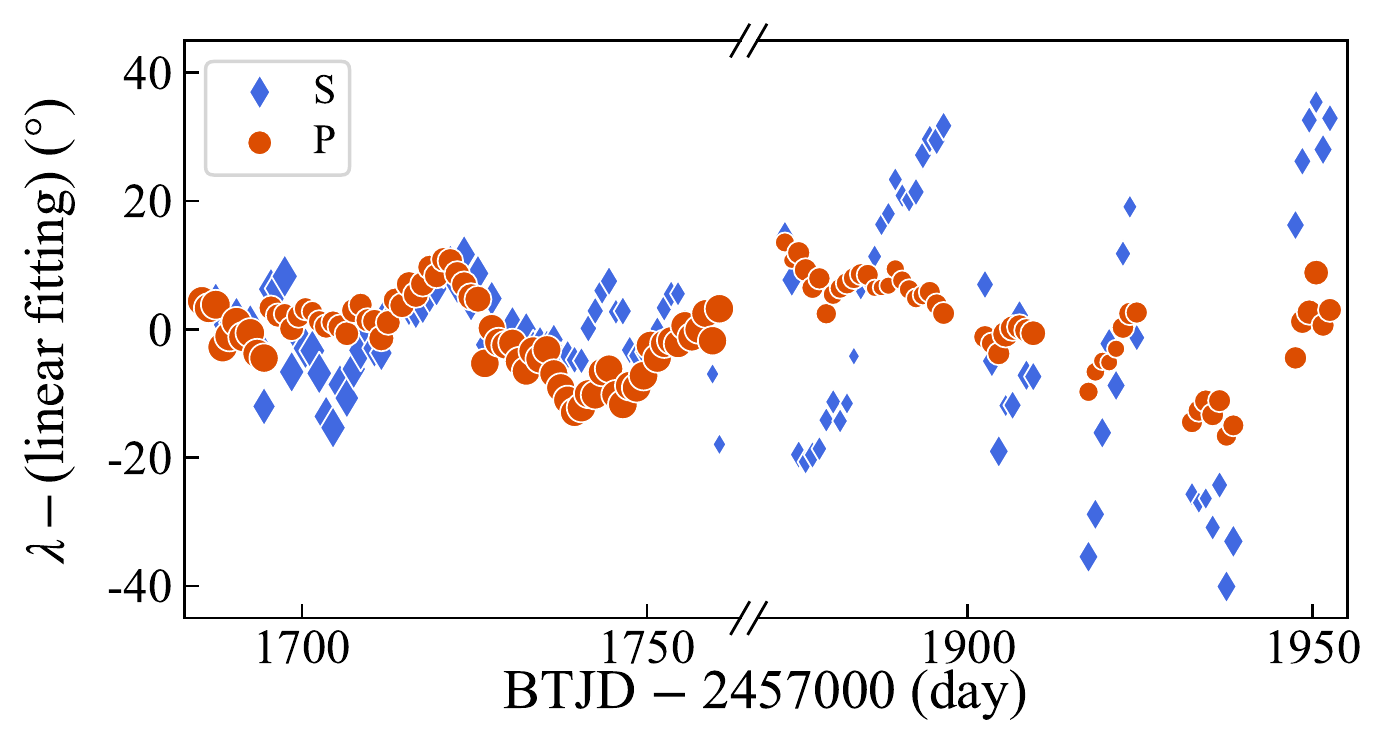}
    \caption{Subtraction of spot longitude ($\lambda$) by its linear fitting, $\lambda - $(linear fitting), i.e. spot longitudinal migration from its average location, versus time. The symbols are scaled proportional to the spot radii, as in figure~\ref{fig:lc_slice}.
    }
	\label{fig:LCM_evolve}
\end{figure}

%The most notable phenomenon in the starspot evolution is the longitudinal oscillation of both starspots with respect to their average rotations.
The deviation of spot longitude from its rotational modulation period represents the spot longitudinal migration from average location, as shown in figure~\ref{fig:LCM_evolve}.
In "T1", the two spots exhibited highly synchronized oscillatory variations with amplitude of about $15\degr$ and period of about $40$ days. This phenomenon can also be seen in figure~\ref{fig:lc_slice_vertical} where the two dips in LC varied periodically with time.
%If not by chance, such a synchronization might be an evidence of some kind of internally physical connection between spots pair corresponding to the magnetic structure inside the stellar sphere.
%The spot migration phenomena were widely discovered on both eclipsing binaries and singles in multiple timescales \citep[e.g.][]{Lanza2006, Huber2010, Xiang2020}. The migration of spot pairs was monitored in long-term studies and found to result in interesting phenomena for example the so-called "flip-flop" \citep{Olah2006, }, however we had not found detailed analysis on the synchronized migration of spots in timescale as ours and the further study of such phenomenon extends the scope of this paper.
In "T2", "P" remained the same level of evolution like before, however "S" underwent a much sudden evolution than "T1", i.e. it had an obviously small average rotational period than "P" and exhibited a much larger longitudinal oscillation with amplitude of about $40\degr$ and period of $30-40$ days, which broke the previous synchronization and seemed to be a new distribution of the spot feature.
This indicates that "P" survived over time spanning from "T1" to "T2", while "S" survived over "T1" but might be a new one in "T2".

On the other hand, although correlated with recovered latitudes as shown in figure~\ref{fig:LCM_rad_lat}, % indicating that it may be misleading and should be treat with caution. %that the variations of spot sizes from best-fit solutions may be misleading and thus should be treated with caution.
the spot sizes were found to evolve with time in terms of relative variations.
%there are evidences of its evolution can be found by comparing the relative difference between spots and estimating the temporal variation of individual spot.
%
Firstly, %the spot sizes are preferred to be different between "T1" and "T2".
compared with a supposed spot located at average latitude $\beta=70\degr$, "P" has a maximal radius of about $28\degr$ in "T1" and decreases to about $20\degr$ in "T2", while "S" has an average radius of about $23\degr$ in "T1" and a smaller radius of only $13\degr$ in "T2".
So the spot configuration was likely to be dominated by one larger plus a smaller active region, and both of them decreased to smaller scales from "T1" to "T2", in the sense of projected area.
Secondly, %deviation of points from the radius-latitude correlation, i.e. simulated lines in figure~\ref{fig:LCM_rad_lat}, indicates reliable evolution of spot size. % comparing to the reference value.
%A sudden increase of "S" and simultaneous decrease of "P" are notable around epoch $1961$, 
as shown in figure~\ref{fig:LCM_comp}, "S" radius in "T1" increased rapidly around epoch $1692$ from $26\degr$ to $31\degr$ and then decreased gradually to small value of about $12\degr$ at the end of "T1", which was in similar size as in "T2", while "P" underwent an opposite variation which had a rapid decrease at epoch $1692$ from $29\degr$ to $22\degr$ followed by a gradually recovery to large size.
%Another notable evolution is the continuous decrease of "S" in size from $28\degr$ at epoch $1730$ to small value of $12\degr$ similar to "T2".
For the case of "T2", in contrary, the two spots showed synchronized variation of radii over time.

Considering the spot radius-latitude correlation, the variations in "T2" might result from the parameter degeneracy due to weak constraints of photometry, however one can still recognize two evidences of spot evolution in "T1" with reliability.
One is the simultaneous but opposite variations between "P" and "S" around $1692$, which indicates either a switch of the activity level from one spot to another or the latitudinal migration to opposite directions. This phenomenon was also reported and analysed in detail on HQ Hydrae by \citet{Lehtinen2012} who discriminated the "flip-flop" from more commonly "switch" phenomenon.
The other is the decrease of "S" since $1720$.

\begin{figure}
	\includegraphics[width=\columnwidth]{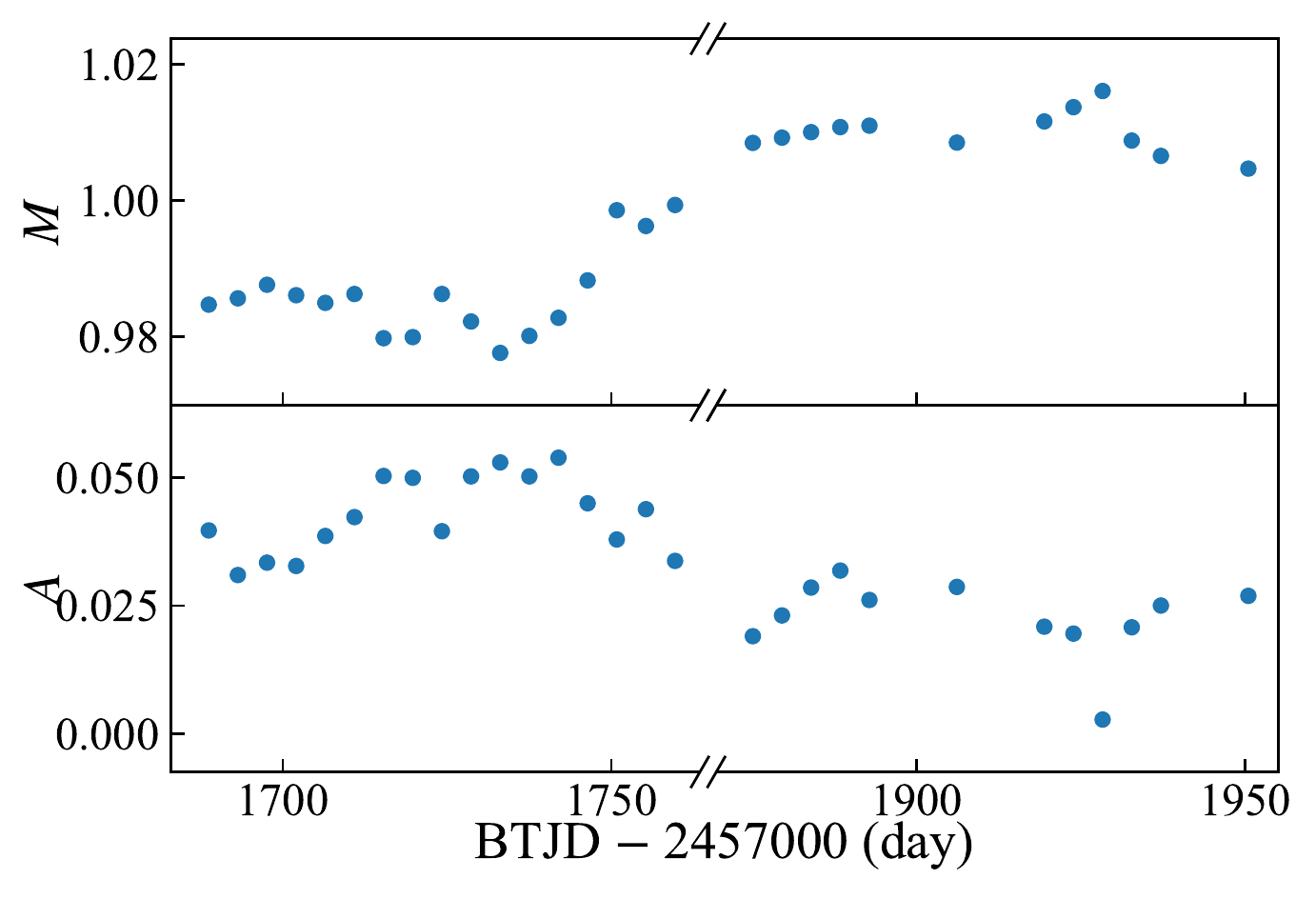}
    \caption{The long term variation of the mean brightness $M$ (top) and peak-to-peak amplitude $A$ (bottom) of HD 134319 measured from figure~\ref{fig:lc_slice_vertical}.}
	\label{fig:lc_A}
\end{figure}

Figure~\ref{fig:lc_A} presents the long-term evolution of the LC mean brightness $M$ and peak-to-peck amplitude $A$, which were usually employed in estimating the stellar activity variations due to spots \citep{Berdyugina2002, Lehtinen2012, Ioannidis2020}. The variation in $M$ could be explained by activity difference, or spot size variations, between "T1" and "T2". While the decrease in $A$ attributes mainly to the decay of spot "P" to small size from "T1" to "T2", supporting the above analysis.

Considering the coherence of the spot longitudinal migration and radial evolution, the primary active region "P" is likely to be the same one which migrated and oscillated from phase $0.71$ and lived longer than the observation season ($280$ days), while the secondary active region "S" exhibited different types of evolution and thus is likely to represent different spot features at least between "T1" and "T2".
This inference is in accordance with the scenario that spot lifetime is proportional to its size and large spot can survive for many years (see reviews of \citet{Berdyugina2005} and \citet{Strassmeier2009}).
%However a quantitative analysis is impossible in our case because no evidence of emergence and disappearance of spots can be found within observation era.

\subsubsection{Differential rotation}\label{sec:SDR}
Despite the weak constraint on spot latitude, the multiple rotational modulation periods in abundant photometric time series were widely taken as indicator of SDR on many stars other than the Sun. By assuming the latitudinal occupation of spots towards both equator and pole as much as possible, the seasonal period variation from extreme long-term photometry was applied to estimate the lower-limit of SDR \citep[e.g.][]{Henry1995, Messina2003, Balona2016}. At the mercy of precise photometric data, there exist studies attempting to inverse spot rotational modulation period and latitude simultaneously and thus directly estimating the SDR \citep[e.g.][]{Croll2006, Frohlich2012, Lanza2016}. However, studies aiming at examining the reliability of SDR detection based on photometry alone revealed the probability of misleading results especially in evolving stages \citep{Aigrain2015, Basri2020}.

The difference of spot rotational modulation periods between "P" and "S" in "T1" is quite small, i.e. $\Delta P = 0.0010310\pm0.0016966\,\mathrm{day}$, corresponding to a lower-limit of SDR as $\alpha = 0.000233$. % using equation~\ref{eq:SDR} under an assumption of one equatorial and simulations one polar spot.
Meanwhile, the difference between "P" and "S" in "T2" yields $\Delta P = 0.0656278\pm0.0064641\,\mathrm{day}$, corresponding to a lower-limit of SDR as $\alpha = 0.0148$, which is apparently smaller than prediction from statistics \citep{Reinhold2013, Balona2016} and dynamo studies \citep{Kuker2008, Kitchatinov2012}.
Such a small difference is probably due to the nearby latitudinal distribution of spots during the observation season (or, spot pair is located on opposite hemispheres, which is indistinguishable in the case of inclination $i=90\degr$).
As in figure~\ref{fig:LCM_comp}, the best-fitted spot latitudes were close to each other, i.e. between $\beta=50\degr$ and $85\degr$ and varied temporally not too much, supporting above situation.

Due to the weak constraint of photometric observation on latitudinal information and the possible migration of starspots, the estimation of SDR is pretty difficult. The lower limit of SDR shear is reliable only when the magnetic configuration can be described by such a two-spot model and all starspots are centered in certain longitudes, however both of which cannot be confirmed from collected data, and thus the above discussion should be treated with caution.
%Whatever, due to the lack of reliable latitude information, it is hard to directly estimate the SDR on HD 134319 in this paper.
%But we estimated the rotational periods of individual spots with reliability.
%We want to note that the measurement of spot rotational period is reliable only when spot is stable. The best-fit solutions reveals stable distribution of both "P" and "S" in time during time $1683-1764$ 

%% ======
\subsection{Chromospheric activity in short and long timescales}
\begin{figure*}
	\includegraphics[width=0.95\textwidth]{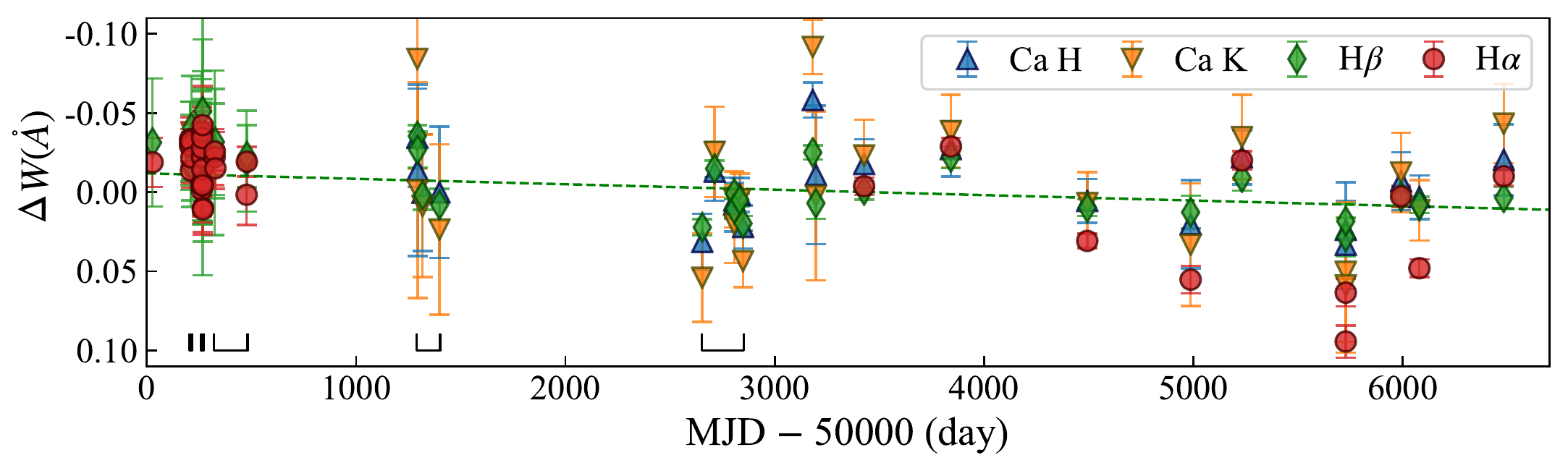}
    \caption{The relative equivalent widths versus time as table~\ref{tab:ew}. The linear fitting on \Hb~is overlaid with dashed line, and the time durations used for analysing rotational modulation (figure~\ref{fig:ew_RotMod}) are marked with dark lines.
    Hereafter, the axis corresponding to equivalent widths is inverted for easier view.
    }
	\label{fig:ew_time}
\end{figure*}

The relative equivalent widths ($\Delta W$s) of \CaII~H and K, \Hb~and \Ha~lines are listed in table~\ref{tab:ew} and plotted in figure~\ref{fig:ew_time}.
% median $\sigma: 0.0360875 (Hb), 0.015585 (Ha) for OHP/ELODIE and 0.0259955 (K), 0.018091 (H), 0.006445 (Hb), 0.0059265 (Ha) for Keck/HIRES
Observations by OHP/ELODIE within about one year have good phase coverage, the measurements yield typical median uncertainties of $0.036088$ for \Hb~and $0.015585$ for \Ha, except two extremely low SNR records 23 and 24.
Observations by Keck/HIRES have long time baseline over 14 years but are sparsely sampled, the measurements yield typical median uncertainties of $0.025996,~0.018091,~0.006445,~\text{and}~0.005927$ for \CaII~H and K, \Hb~and \Ha, respectively.
%The time baseline includes 4 durations within which the observations have good phase coverage. %connection between chromospheric activity variation and the spot rotational modulation may exists.

\begin{table*}
    \caption{The relative equivalent widths measured from the residual spectra, which represent the difference between individual spectrum and the overall spectrum for each chromospheric activity indicator (see section~\ref{sec:model_ew}). The last colume is the instrument, "E" represents OHP/ELODIE and "H" represents Keck/HIRES.}
    \label{tab:ew}
    \begin{tabular}{rrrrrrrrrrrl}
        \hline
        %No. & Date & epoch &  $\Delta W_{\text{K}}$ & $\sigma_{\text{K}}$ & $\Delta W_{\text{H}}$ & $\sigma_{\text{H}}$ & $\Delta W_{\text{H}\beta}$ & $\sigma_{\text{H}\beta}$ & $\Delta W_{\text{H}\alpha}$ & $\sigma_{\text{H}\alpha}$ & Instrument \\
        %  & yyyymmdd & MJD-50000 & ($\mathring{A}$) & ($\mathring{A}$) & ($\mathring{A}$) & ($\mathring{A}$) & ($\mathring{A}$) & ($\mathring{A}$) & ($\mathring{A}$) & ($\mathring{A}$) & \\
        No. & Date & Epoch &  $\Delta W_{\text{K}}$ & $\Delta W_{\text{H}}$ & $\Delta W_{\text{H}\beta}$ & $\Delta W_{\text{H}\alpha}$ & Inst. \\
            & yyyymmdd & MJD-50000 & ($\mathring{A}$) & ($\mathring{A}$) & ($\mathring{A}$) & ($\mathring{A}$) & \\
        \hline
         1&19951106&  27.758241&...&...&-0.031373$\pm$0.040562&-0.018857$\pm$0.015511&E\\
         2&19960503& 206.975949&...&...&-0.029332$\pm$0.027837&-0.033498$\pm$0.013834&E\\
         3&19960503& 206.989294&...&...&-0.030835$\pm$0.026398&-0.030377$\pm$0.013863&E\\
         4&19960504& 208.035718&...&...&-0.027873$\pm$0.021146&-0.029040$\pm$0.010667&E\\
         5&19960505& 209.994572&...&...&-0.015346$\pm$0.020592&-0.032421$\pm$0.010319&E\\
         6&19960506& 210.940266&...&...&-0.017242$\pm$0.026576&-0.013584$\pm$0.010882&E\\      % manually modify MJD from 209.940266 to 210.940266 wrt OBSTIME.
         7&19960508& 212.998855&...&...&-0.040752$\pm$0.032760&-0.021943$\pm$0.016166&E\\
         8&19960628& 262.953877&...&...&-0.018798$\pm$0.036594&-0.001461$\pm$0.016825&E\\
         9&19960628& 262.962755&...&...&-0.025168$\pm$0.039243&-0.004372$\pm$0.017009&E\\
        10&19960629& 263.881146&...&...&-0.035705$\pm$0.035581&-0.022479$\pm$0.016555&E\\
        11&19960629& 263.890013&...&...&-0.039269$\pm$0.037212&-0.027851$\pm$0.015313&E\\
        12&19960630& 264.888473&...&...&-0.030101$\pm$0.033388&-0.038060$\pm$0.015731&E\\
        13&19960630& 264.897384&...&...&-0.018138$\pm$0.038456&-0.034214$\pm$0.016794&E\\
        14&19960701& 265.879850&...&...&-0.011007$\pm$0.031238&-0.006143$\pm$0.014964&E\\
        15&19960701& 265.892223&...&...&-0.020166$\pm$0.038616&-0.014393$\pm$0.017049&E\\
        16&19960702& 266.930995&...&...&-0.051228$\pm$0.070634&-0.042379$\pm$0.024720&E\\      % large err
        17&19960702& 266.939896&...&...&-0.021975$\pm$0.074569&-0.004292$\pm$0.023377&E\\      % large err
        18&19960703& 267.924133&...&...&-0.006526$\pm$0.037805& 0.009826$\pm$0.015257&E\\
        19&19960703& 267.932987&...&...&-0.023558$\pm$0.042209& 0.010922$\pm$0.015659&E\\
        20&19960828& 323.834537&...&...&-0.024840$\pm$0.051999&-0.022079$\pm$0.017712&E\\
        21&19960829& 324.823485&...&...&-0.026197$\pm$0.029949&-0.025716$\pm$0.012245&E\\
        22&19960901& 327.855637&...&...&-0.031695$\pm$0.033460&-0.015077$\pm$0.013060&E\\
        23&19961229& 447.207779&...&...&-0.197406$\pm$0.339053& 0.181427$\pm$0.068684&E\\      % terrible err due to extremely low SNR
        24&19970125& 474.192547&...&...&-0.128411$\pm$0.360773& 0.169545$\pm$0.141722&E\\      % terrible err due to extremely low SNR
        %23&19961229& 447.207779&...&...&...&...&E\\      % terrible err due to extremely low SNR
        %24&19970125& 474.192547&...&...&...&...&E\\      % terrible err due to extremely low SNR
        25&19970128& 477.205753&...&...&-0.019608$\pm$0.031867& 0.001652$\pm$0.019156&E\\
        26&19970129& 478.191632&...&...&-0.022774$\pm$0.019649&-0.019219$\pm$0.009338&E\\
        27&19990424&1292.561160&-0.084215$\pm$0.069323&-0.034360$\pm$0.031120&-0.035413$\pm$0.007033&...&H\\
        28&19990425&1293.526583&-0.001237$\pm$0.068204&-0.013785$\pm$0.054094&-0.026791$\pm$0.011333&...&H\\
        29&19990519&1317.430287& 0.008699$\pm$0.045056& 0.000250$\pm$0.036867& 0.001960$\pm$0.009238&...&H\\
        30&19990808&1398.261886& 0.023564$\pm$0.053873& 0.000142$\pm$0.041517& 0.007955$\pm$0.010172&...&H\\
        31&20030114&2653.679294& 0.054104$\pm$0.027859& 0.031481$\pm$0.017816& 0.022237$\pm$0.005083&...&H\\
        32&20030314&2712.478795&-0.025440$\pm$0.028476&-0.012997$\pm$0.018487&-0.014811$\pm$0.005179&...&H\\
        33&20030616&2806.324191& 0.004558$\pm$0.017696& 0.005198$\pm$0.010904& 0.000763$\pm$0.003301&...&H\\
        34&20030616&2806.375226& 0.020166$\pm$0.024592& 0.007891$\pm$0.017010& 0.010546$\pm$0.006181&...&H\\
        35&20030713&2833.330383& 0.004602$\pm$0.016301& 0.001800$\pm$0.010753& 0.005014$\pm$0.004157&...&H\\
        36&20030728&2848.266842& 0.043567$\pm$0.016419& 0.022013$\pm$0.013506& 0.019795$\pm$0.004684&...&H\\
        37&20040625&3181.446408&-0.091720$\pm$0.017171&-0.058185$\pm$0.011094&-0.025069$\pm$0.004413&...&H\\
        38&20040710&3196.314068& 0.002408$\pm$0.053364&-0.010890$\pm$0.043795& 0.006985$\pm$0.009848&...&H\\
        39&20050225&3426.617754&-0.023167$\pm$0.022615&-0.017766$\pm$0.015543&-0.001114$\pm$0.005827&-0.003765$\pm$0.005488&H\\
        40&20060416&3841.393636&-0.038543$\pm$0.022988&-0.027235$\pm$0.017408&-0.021964$\pm$0.006786&-0.028862$\pm$0.005404&H\\
        41&20080127&4492.683551& 0.006933$\pm$0.019474& 0.005565$\pm$0.013863& 0.009798$\pm$0.005215& 0.030829$\pm$0.004608&H\\
        42&20090604&4986.408533& 0.033238$\pm$0.038755& 0.020317$\pm$0.027897& 0.012719$\pm$0.010399& 0.055268$\pm$0.008783&H\\
        43&20100204&5231.671857&-0.034793$\pm$0.026827&-0.022060$\pm$0.017126&-0.009158$\pm$0.008149&-0.020032$\pm$0.005889&H\\
        44&20110615&5727.448702& 0.050378$\pm$0.044145& 0.023823$\pm$0.030052& 0.028416$\pm$0.012267& 0.094395$\pm$0.010165&H\\
        45&20110615&5727.458635& 0.058932$\pm$0.042421& 0.033285$\pm$0.027970& 0.018301$\pm$0.010821& 0.063508$\pm$0.008704&H\\
        46&20120305&5991.668097&-0.012412$\pm$0.025164&-0.006934$\pm$0.018366& 0.002887$\pm$0.006191& 0.002646$\pm$0.006686&H\\
        47&20120601&6079.331947& 0.011411$\pm$0.019098& 0.003193$\pm$0.013857& 0.008064$\pm$0.005063& 0.048057$\pm$0.005655&H\\
        48&20130709&6482.252854&-0.043213$\pm$0.024948&-0.020351$\pm$0.022389&-0.003531$\pm$0.006699&-0.010131$\pm$0.005964&H\\
        \hline
    \end{tabular}
\end{table*}

%\textcolor{red}{\sout{By using subsets of Keck/HIRES spectra, \citet{Wright2004, Isaacson2010} and \citet{Butler2017} measured S-index of HD 134319 independently. Comparison between our measurements and their results is shown in figure~\ref{fig:ew_comp}, overlaid with respective linear fittings. Note that \citet{Wright2004} employed a scheme similar to us by using the highest SNR observation as a template in measuring the "sensitive differential S-values". Our result fits well with the ones of \citet{Wright2004} and \citet{Isaacson2010}, indicating the reliability of our scheme in measuring the relative variation of chromospheric activity indicators. }}

\subsubsection{Long-term evolution}
As shown in figure~\ref{fig:ew_time}, during observations in near 20 years, the chromospheric activity indicators show variations exceed $0.1\mathring{A}$.
%A quasi-periodic variation of their strengths in long timescale was visible but with low confidence due to the poor phase coverage.
Evolutions in both short and long timescale are discriminable by comparing their relative variations of equivalent widths in different timescales.
The short-term evolution within one or a few rotations can be attributed to the rotational modulation as revealed by OHP/ELODIE observation with good phase coverage, 
which will be discussed in more detail in section~\ref{sec:ew_rot}.
While the notable variation in Keck/HIRES observing season shows the maximal scale, indicating evolution in long timescale.
Linear fitting of \Hb~(overlaid by green dashed line) reveals a trend of its average $\Delta W$ from $-0.012$ to $0.011\mathring{A}$ with time, implying a decrease of activity level in long timescale.

\subsubsection{Correlations between indicators}
Correlations between chromospheric activity indicators were widely studied on active stars \citep[e.g.][]{Strassmeier1990, Montes1995, Cincunegui2007, Martinez-Arnaiz2011, Scandariato2017}.
The excess emissions of \CaII~H and K ($3968$ and $3933\,\mathring{A}$) resonance were found to be tightly correlated with each other and thus were practically measured together, for example, the widely used S-index \citep{Duncan1991}, in estimating the stellar activity.
Two other well studied chromospheric proxies, \Ha~($6363\,\mathring{A}$) and \Hb~($4861\,\mathring{A}$) lines, which can be feasibly measured with high SNR on even pretty cool stars where the \CaII~resonance becomes progressively faint, resulted from the two most probable transitions of electrons between energy levels of Hydrogen which is the most abundant element inside the Sun and other stars, were also found to be correlated with each other.
However, correlation between \Ha~and \CaII~resonance was more complicated, because \Ha~reveals not only a non-monotonically variation with increasing heating rate \citep{Linsky2017} but also the dependence on stellar physical parameters such as pressure and temperature \citep{Cincunegui2007}.

\begin{figure}
	\includegraphics[width=\columnwidth]{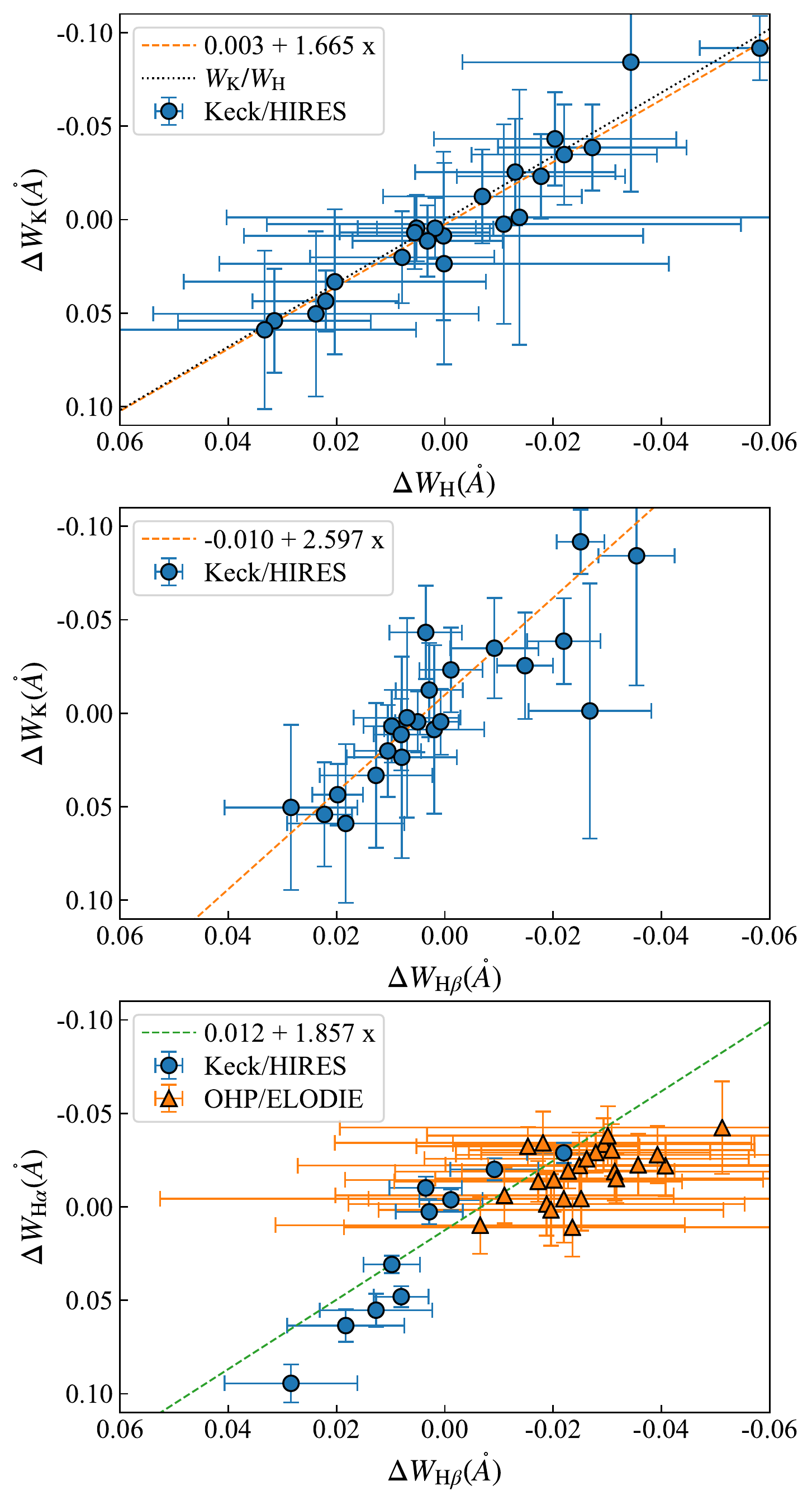}
    \caption{Relative equivalent widths of \CaII~K versus H (top), \CaII~K versus \Hb~(middle) and \Ha~versus \Hb~(bottom), the uncertainties are plotted by horizontal and vertical errorbars, respectively. Weighted linear fittings are overlaid with dashed lines, the ratio of $W_\text{K}/W_\text{H} \sim 1.698487$ is overlaid with dark dots.
    %Note that OHP/ELODIE has much larger uncertainty and thus plays much smaller role than Keck/HIRES in fitting (bottom).
    }
	\label{fig:ew_cor}
\end{figure}

Figure~\ref{fig:ew_cor} shows the correlation between $\Delta W$s of \CaII~K and H, \Hb~and \Ha~lines. Linear fitting to each relation was done for estimating the relative difference between indicators and overlaid in this figure by dashed line. Note by the linear fitting a positive slope yields a positive correlation. % and the intercept represents the difference between respective overall spectra.
We can find that the relative variation of any indicator is strongly correlated with the others, despite the quantitatively different uncertainty levels.

Due to the lack of comparing star as template, it is difficult to estimate the absolute value of \Hb~and \Ha~emissions, while we can estimate equivalent widths of \CaII~H and K lines due to their narrow emission profiles. Measurements gave $W_\text{H} = -0.342627\,\mathring{A}$ and $W_\text{K} = -0.581948\,\mathring{A}$ for \CaII~H and K lines, respectively, by directly integrating their overall spectra over $1.1\,\mathring{A}$ width portions centred on the emission cores, and yielded a ratio of $W_\text{K}/W_\text{H} = 1.698487$ which is close to the slope of linear fitting (top panel in figure~\ref{fig:ew_cor}). This infers that the relative variations of \CaII~H and K lines are proportional to their respective absolute emission strengths.

The larger variation of \Ha~than \Hb~in bottom panel of figure~\ref{fig:ew_cor} could be attributed to its lower energy required in electron transitions in principle, while the more commonly mentioned correlation between \CaII~and \Ha~lines was found to depend on stellar parameters such as the stellar effective temperature, metallicity and pressure etc. \citep[e.g.][]{Cincunegui2007, Walkowicz2009, Scandariato2017}.
By analogy to the study on the Sun by \citet{Gebbie1974}, who proposed that the formation of \CaII~is direct collision dominated due to turbulent velocities while the formation of \Ha~is photoionization dominate due to radiation which decreases in late-type stars, one could reasonably expect different correlations between \CaII~and \Ha~on different type of stars.
Thus the ratio between \CaII~and \Ha~ (simply as $(\Delta W_\mathrm{K}+\Delta W_\mathrm{H})/\Delta W_{\mathrm{H}\alpha} \sim 2.238$) is larger than one, might indicate that the chromospheric excitation on HD 134319 is dominated by collision due to turbulent velocities.
% (H+K)/Ha = H/K*K/Hb*Hb/Ha + K/Hb*Hb/Ha = (1/1.665 + 1) * 2.597 * 1/1.857 = 2.238

\subsubsection{Rotational modulation of indicators}\label{sec:ew_rot}
\begin{figure}
    \includegraphics[width=\columnwidth]{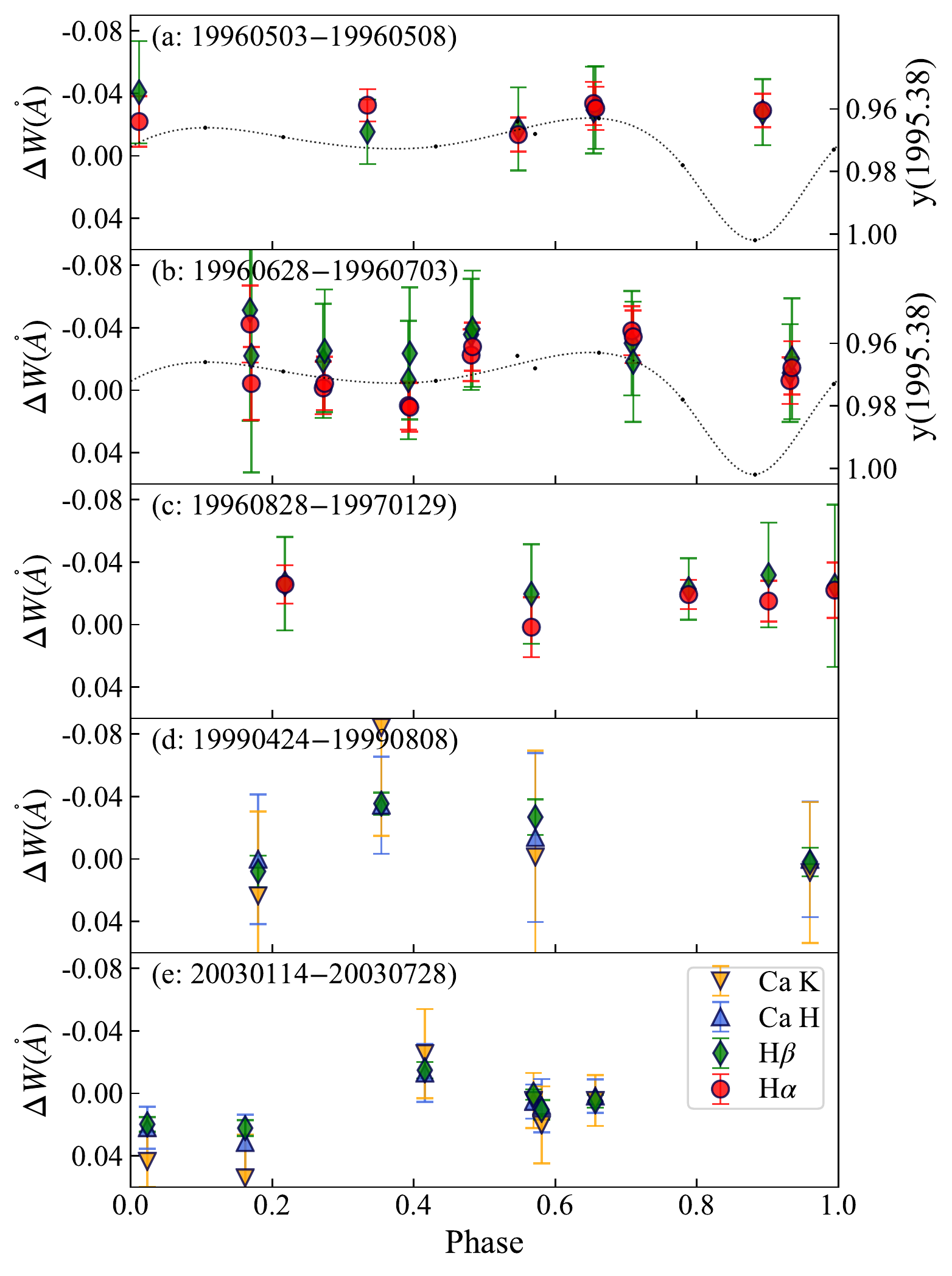}
    \caption{Phase--folded relative equivalent widths ($\Delta W$s) of \CaII~K and H, \Hb~and \Ha~lines in five chosen time durations shorter than half a year (marked in figure~\ref{fig:ew_time}).
    Peaks in the phased curves represent the phase of enhanced chromospheric emissions under rotational modulation.
    Moreover, the y-band LC taken from figure 3 of \citet{Messina1998} is re-folded by newly measured rotation period and plotted in (a) and (b) with dark dots, overlaid by smoothed dotted lines, and is inverted so that the possible spatial connection between chromospheric emissions and photospheric spots is manifested as the positive correlation between inverted $\Delta W$ and LC.
    }
    \label{fig:ew_RotMod}
\end{figure}

% ===
In short timescale, variations of chromospheric proxies under rotational modulation were widely reported on stars \citep[e.g.][]{Wright2004, Isaacson2010, FloresSoriano2017}. Such variations can also be found on HD 134319.
%Although it is impossible to explore the probability of such a connection on HD 134319 due to the lack of overlap between photometric and spectroscopic observations, the variations of chromospheric activity indicators under rotational modulation in several time durations are notable.
%
Figure~\ref{fig:ew_RotMod} shows the phase folded $\Delta W$s of \CaII~H and K, \Hb~and \Ha~within chosen time durations no longer than half a year. Among the five durations, (a) and (b) have good phase coverage in about one rotation cycle, while (c) -- (f) are phase--folded for data collected in months.
With the good phase coverage of OHP/ELODIE observations, two peas around phases $0.2$ and $0.7$ are recognizable from panels (a) -- (b) during time from May to July of 1996, corresponding to two active regions characterized by enhancement of chromospheric emissions.
Later, for the precise data observed by Keck/HIRES in year 1999 and 2003, at least one active region around phase $0.4$ can be recognized, despite the poor phase coverage.

The long-term behaviour of the chromospheric emissions under rotational modulation implies that HD 134319 was dominated by two active regions during years from 1996 to 1997, and the pea around phase $0.2-0.4$ might be due to a long-lived active region lasting over years from 1996 to 2003.

\subsubsection{Connection between chromospheric activity and photospheric spots}
The possible connection between chromospheric activity and photometric variation would indicate a configuration of the magnetic field spreading over photosphere and chromosphere. A clear anti-correlation between them on LQ Hydrae was reported and interpreted as the spatial connection between photospheric dark spots and chromospheric plages \citep{CaoDT2014, FloresSoriano2017}, the similar phenomenon was also reported on FK Comae \citep{Vida2015}.

It is difficult to analyse this relation exactly on HD 134319 due to the lack of overlap between photometric and spectroscopic observations analyzed in this paper.
%\textcolor{red}{\sout{However, considering the photometric analyses of \citet{Messina1998}, \citet{Messina1998b} and section~\ref{sec:spot_evolve_DR}, starspots on HD 134319 are probably to survive over half a year. We can obtain some approximate results by comparing spectroscopic data by OHP/ELODIE in 1996 with photometric observation by \citet{Messina1998} in 1995, i.e. separated by one year.}}
%It is reasonable to assume that the chromospheric activity indicators under rotational modulation should vary not too much if the connection between the chromospheric activity and photospheric spots exists on HD 134319.
In figure~\ref{fig:ew_RotMod} (a) and (b), we over-plot the phased LC taken from figure 3 of \citet{Messina1998} observed around 1995.38, which is the closest to the spectroscopic data in time, seperated by more than one year. A positive correlation between $\Delta W$ and inverted photometric LC can be found but at low significance. Thus no reliable conclusions can be concluded with collected data.
%\textcolor{red}{\sout{Although the spot configuration might evolve notably in timescale over one year and result in apparently different LC, as intuitively found in \citet{Messina1998} and revealed by the difference of "S" between "T1" and "T2" of TESS (section~\ref{sec:spot_evolve}), one can still see that the $\Delta W$ varies qualitatively in positive correlation with the inverted photometric LC. This, if not by chance, indicates the spatial connection between the chromospheric emissions and photospheric spots, as observed on above mentioned LQ Hyadrae and FK Comae.}}

%One can thus expect that the two-spot configuration, which was preliminarily reported during years from 1991 to 1995 by \citet{Messina1998} and recently observed by TESS with precision, may frequently emerge over tens of years, and there also may exist long-lasting spot surviving over years on HD 134319.

%% file: 5-conclusion.tex
\section{Summary}\label{sec:summary}

In this paper, we present the analysis of the starspot configuration, evolution and chromospheric activity on HD 134319, by employing the high precision photometry by space-based TESS telescope in sectors 14--16 (epoch 1683--1764, called "T1") and 21--23 (epoch 1869--1955, "T2") and the spectroscopic data observed by OHP/ELODIE and Keck/HIRES from year 1995 to 2013.

We firstly measured the rotation period of the star from TESS LC as $P=4.436391\pm0.00137$ days which is used in subsequent analysis, using GLS method \citep{Zechmeister2009}.
Then, the starspot configurations at series of epochs were derived by splitting the whole TESS LC into 129 chunks in $5$ days length and then fitting them separately with a two-spot model.
Besides, based on analysis of the high-resolution spectroscopic data, we derived the relative equivalent widths ($\Delta W$s) of the \CaII~H and K, \Hb~and \Ha~lines to investigate its chromospheric activity.

We focus on magnetic characteristics on HD 134319 using a simple two-spot model. One should keep in mind that the actual number of starspots (or active regions) could not be decided with collected data and might much larger than the number of dips in LC. However our model revealed reasonable good fits, and if the star really has only two spots or can be described by such a model, it would be the result.
Our main results are summarized as follows:

(\UpperRoman{1})
As revealed by %\textcolor{red}{\sout{joint}} 
analysis of %\textcolor{red}{\sout{photometric and spectroscopic} 
TESS LC data, a two-spot configuration, i.e., a long-lasting primary spot "P" plus a secondary spot "S", %\textcolor{red}{\sout{was likely to frequently emerge over tens of years}
was capable of explaining the LC variation on HD 134319 during observation. Furthermore, the primary spot was likely to survive at least longer than TESS observation duration, i.e. 300 days, and might survive over years.

(\UpperRoman{2})
Linear fittings on spot longitudes give the average rotational modulation period of spots as: $P_\mathrm{P} = 4.4236654\pm0.0003350\,\mathrm{day}$ emerging at phase $\lambda_\mathrm{P}=0.71$ %corresponding to
at
 epoch 1683, $P_\mathrm{S1} = 4.4226344\pm0.0016632\,\mathrm{day}$ emerging at phase $\lambda_\mathrm{S1}=0.34$ %in
at epoch 1683 and $P_\mathrm{S2} = 4.3580376\pm0.0064554\,\mathrm{day}$ emerging at phase $\lambda_\mathrm{S2}=0.19$ %in
at epoch 1689.
This corresponds to a low-limit of SDR, if starspots underwent no migration at long-time scale, as $\alpha = 0.0148$ which is weaker than prediction, %and thus
which might due to a nearby distribution of spot latitudes. %\textcolor{red}{But this conclusion is excluded by assuming that starspots underwent no longitudinal migration.}
% the best-fit latitidues from spot modelling indicate that the spots are likely to centered on close latitudes.

(\UpperRoman{3})
%Notable evolution of spot configuration was revealed by spot modelling within TESS era.
%
Reliable evolutions of spot radii can be derived despite the radius-latitude degeneracy (figures~\ref{fig:LCM_comp} and \ref{fig:LCM_rad_lat}). A sudden increase of "S" from radius $\gamma=26\degr$ to $31\degr$ and simultaneous sudden decrease of "P" from $\gamma=29\degr$ to $22\degr$ around epoch $1692$ indicate an exchange of activity strength between spots.
Since epoch $1692$, "S" radius underwent a gradual decrease to about $12\degr$ at the end of "T1", similar to its radius in "T2".
Besides, decrease of spot radii from epoch $1683-1764$ to $1869-1955$ was notable (figure~\ref{fig:LCM_rad_lat}). Comparing with a reference spot at latitude $\beta=70\degr$, "P" had radius $\gamma\sim28\degr$ and "S" had radius $\gamma\sim23$ in epoch $1683-1764$, while "P" had radius $\gamma\sim20\degr$ and "S" had radius $\gamma\sim13\degr$ in epoch $1869-1955$. This indicates the decrease of magnetic activity from epoch $1683-1764$ to $1869-1955$.

(\UpperRoman{4})
The spots also exhibited longitudinal migrations. "P" evolved slowly and migrated in oscillation around its average longitude with amplitude of about $15\degr$ and period of about $40$ days, "S" in "T1" showed migration tightly synchronized with "P", while "S" in "T2" showed much larger oscillation with amplitude of about $40\degr$ and period of $30-40$ days (figures~\ref{fig:lc_slice} and \ref{fig:LCM_evolve}).

(\UpperRoman{5})
Relative variations of chromospheric activity indicators from 1995 to 2013 revealed both short-term rotational modulation (figure~\ref{fig:ew_RotMod}) and long-term decrease of activity (figure~\ref{fig:ew_time}), impling the existence and evolution of magnetic activity, and thus the distortion of LC is likely due to starspots on HD 134319.
%\textcolor{red}{\sout{Their coincidence with photometric light curve (figure~\ref{fig:ew_RotMod}) implies a possible spatial connection between chromospheric emissions and photospheric starspots on HD 134319.}}

%% file: 6_Acknowledgement.tex
\section*{Acknowledgments}
%We would like to thank the staff of the 2.4-m telescope of Yunnan Observatories for supporting our observations.
%Funding for the 2.4 m telescope has been provided by the Chinese Academy of Sciences and the People's Government of Yunnan Province.
This research has made use of the SIMBAD database, operated at CDS, Strasbourg, France (\url{http://cdsweb.u-strasbg.fr/}),
the data from the European Space Agency (ESA) mission {\it Gaia} (\url{https://www.cosmos.esa.int/gaia}), processed by the {\it Gaia} Data Processing and Analysis Consortium (DPAC, \url{https://www.cosmos.esa.int/web/gaia/dpac/consortium}),
the Keck Observatory Archive (KOA), which is operated by the W. M. Keck Observatory and the NASA Exoplanet Science Institute (NExScI), under contract with the National Aeronautics and Space Administration
and the ELODIE archive at Observatoire de Haute-Provence (OHP, \url{http://atlas.obs-hp.fr/elodie/}).
Funding for the DPAC has been provided by national institutions, in particular the institutions participating in the {\it Gaia} Multilateral Agreement.
%
% SIMBAD
%This research has made use of the SIMBAD database, operated at CDS, Strasbourg, France (\url{http://cdsweb.u-strasbg.fr/}),
% Gaia
%the data from the European Space Agency (ESA) mission {\it Gaia} (\url{https://www.cosmos.esa.int/gaia}), processed by the {\it Gaia} Data Processing and Analysis Consortium (DPAC, \url{https://www.cosmos.esa.int/web/gaia/dpac/consortium}). Funding for the DPAC has been provided by national institutions, in particular the institutions participating in the {\it Gaia} Multilateral Agreement.
% KOA official acknowledgement.
%This research has also made use of the Keck Observatory Archive (KOA), which is operated by the W. M. Keck Observatory and the NASA Exoplanet Science Institute (NExScI), under contract with the National Aeronautics and Space Administration, %.
% ELODIE
%Based on spectral data retrieved from
%and the ELODIE archive at Observatoire de Haute-Provence (OHP, \url{http://atlas.obs-hp.fr/elodie/}).
%
TESS photometric data presented in this paper were obtained from the Mikulsky Archive for Space Telescopes (MAST). STScI is operated by the Association of Universities for Research in Astronomy, Inc., under NASA contract NAS5-26555. 
%This paper includes data collected by the Kepler mission. Funding for the Kepler mission is provided by the NASA Science Mission directorate.
This paper includes data collected by the TESS mission. Funding for the TESS mission is provided by the NASA's Science Mission Directorate.
% =======================
% LightKurve
We made use of {\sc Lightkurve}, a Python package for Kepler and TESS data analysis \citep{Lightkurve:2018} launched in Jupyter Notebook environment (\url{https://github.com/jupyter/notebook/}), built on top of libraries including {\sc NumPy} \citep{numpy:2020}, {\sc SciPy} \citep{scipy:2020} and {\sc Matplotlib} \citep{matplotlib:2007} and relative to {\sc astropy} \citep{astropy:2018}, {\sc astroquery} \citep{astroquery:2019}, {\sc celerite} \citep{celerite:2017} and {\sc tesscut} \citep{tesscut:2019}.
{\sc NumPy}, {\sc Matplotlib}, {\sc SciPy} and {\sc brokenaxes} (\url{https://github.com/jerrylikerice/brokenaxes}) were used in preparing figures of this paper.
We also thank Dr. Longcheng Gui$^{ \href{https://orcid.org/0000-0001-7766-4815}{\includegraphics[width=5px]{ORCIDiD_icon128x128.png}}}$
%(\href{mailto:guilongcheng@hunnu.edu.cn}{guilongcheng@hunnu.edu.cn}, Department of Physics, Hunan Normal University, Changsha, China) 
for his great help in computing resources.
This work is supported by National Natural Science Foundation of China through grants 
Nos. 10373023, 10773027, 11333006, U1531121, and 11903074. We acknowledge the science research grant from the China Manned Space Project with NO. CMS-CSST-2021-B07.
The joint research project between Yunnan Observatories and Hamburg Observatory is funded by Sino-German Center for Research Promotion (GZ1419).
% This paper made use of iPython (Péreza & Granger 2007), and the python libraries numpy, matplotlib, matplotlib_venn, and asciitable.
%The joint research project between Yunnan Observatories and Hamburg Observatory is funded by Sino-German Center for Research Promotion (GZ1419).
%

\section*{Data Availability}
The data underlying this article are available in the article. % and in its online supplementary material.
Part of the data underlying this paper are in the public domain and available in: the Gaia Archive (\href{https://gea.esac.esa.int/archive/}{https://gea.esac.esa.int/archive/}), the Keck Observatory Archive (KOA; \href{https://koa.ipac.caltech.edu/}{https://koa.ipac.caltech.edu/}), the ELODIE Archive (\href{http://atlas.obs-hp.fr/elodie/}{http://atlas.obs-hp.fr/elodie/}) and Mikulski Archive for Space Telescopes (MAST; \href{https://archive.stsci.edu}{https://archive.stsci.edu}).
%Additional data underlying this paper will be shared on reasonable request to the corresponding author.